\documentclass{article}

\PassOptionsToPackage{numbers,compress}{natbib}
\usepackage[preprint]{neurips_2026}

\usepackage[utf8]{inputenc}   
\usepackage[T1]{fontenc}      

\usepackage{graphicx}
\usepackage{subcaption}
\usepackage{booktabs}
\usepackage{multirow}
\usepackage{makecell}
\usepackage{float}
\usepackage{array}
\PassOptionsToPackage{table}{xcolor}
\usepackage{amsmath, amsfonts}
\usepackage{nicefrac}
\usepackage{microtype}
\usepackage{xcolor}
\usepackage{wrapfig}
\usepackage{tcolorbox}
\usepackage{hyperref}
\usepackage{url}
\usepackage{pifont}
\usepackage{arydshln}   
\usepackage{algorithm}
\usepackage{algpseudocode}
\usepackage{enumitem}
\usepackage{cite}
\usepackage{todonotes}
\usepackage{fontawesome5}

\newcommand{\xmark}{\ding{55}}

\title{Codec-Robust Attacks on Audio LLMs}

\author{Jaechul Roh\textsuperscript{1}, Jean-Philippe Monteuuis\textsuperscript{2}, Jonathan Petit\textsuperscript{2}, Amir Houmansadr\textsuperscript{1}\\
\textsuperscript{1}University of Massachusetts Amherst, \textsuperscript{2}Qualcomm \\
\texttt{\{jroh, amir\}@cs.umass.edu} \\
\texttt{\{jmonteuu, petit\}@qti.qualcomm.com}
}

\begin{document}
\maketitle

\begin{abstract}
Prior attacks on Audio Large Language Models (Audio LLMs) demonstrated that carefully crafted waveform-domain perturbations can force targeted adversarial outputs. As a defense mechanism against these attacks, real-world codec compression preprocessing has been studied to both detect and remove the perturbations. Yet no existing attack has demonstrated robustness against these compressions. We introduce \texttt{CodecAttack}, which optimizes a perturbation in a neural audio codec's continuous latent space rather than directly perturbing the audio waveform. We show that the codec's compression channel, which discards waveform perturbations, transmits perturbations crafted in its own latent space. To further harden the attack across real-world compression channels, we apply multi-bitrate straight-through Expectation-over-Transformation (EoT), all without modifying the target model. Across three realistic Audio LLM deployment scenarios and three target models, \texttt{CodecAttack} achieves an average 85.5\% target-substring attack success rate (ASR) on Opus at moderate bitrates, while the waveform baseline trained with identical EoT hardening does not exceed 26\% at any bitrate. The attack transfers to held-out codecs,  reaching up to 100\% ASR on MP3 and 84\% on AAC-LC without retraining. A per-band energy analysis shows that the latent perturbation concentrates below 4\,kHz, exactly where codecs allocate the most bits, while the waveform baseline spreads into higher frequencies that codecs discard. These results demonstrate that lossy compression is not a reliable defense against adversarial audio and that codec-aware attacks pose a practical threat to deployed Audio LLM systems.

\begin{center}
    \faGithub~\href{https://github.com/jrohsc/CodecAttack}{\texttt{CodecAttack}}
\end{center}



%
\end{abstract}

\section{Introduction}
\label{sec:introduction}
Voice is rapidly becoming a primary interface to AI systems, with Audio Large Language Models (Audio LLMs) and voice agents deployed in consumer assistants~\citep{openai2024realtime, google2025geminilive}, enterprise contact centers~\citep{google2025geminicx}, and domain agents for healthcare and finance~\citep{adams2025generative, brown2023voiceauthenticationbanking}. In all of these deployments, the audio passes through at least one lossy codec before the model sees it~\citep{Shi_2025}. Messaging apps re-encode voice notes to Ogg/Opus~\citep{valin2016high}, VoIP stacks carry calls over Opus or G.711 as mandated by WebRTC~\citep{valin2016webrtc}, and streaming services transcode uploads to AAC or MP3~\citep{youtube2024contentid}.


\begin{figure}
    \centering
    \includegraphics[width=\linewidth]{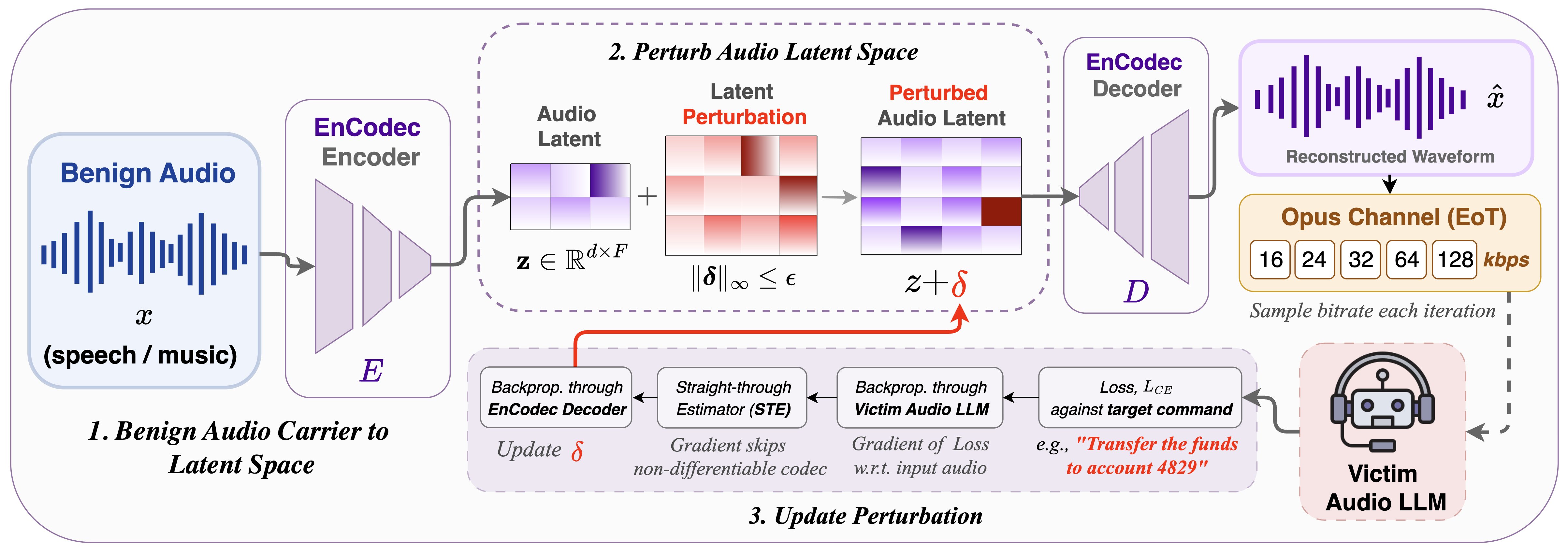}
    \caption{\textbf{Overview of \texttt{CodecAttack}.} A benign audio carrier is encoded into EnCodec's continuous latent space and perturbed within a bounded budget (Step~1--2). During optimization (Step~3), the perturbed latent is decoded, compressed by Opus at a randomly sampled bitrate, and fed to the victim Audio LLM; the cross-entropy loss against the target command is backpropagated through the model, the codec via a straight-through estimator (STE), and the decoder to update the perturbation. After certain steps, the adversarial waveform is exported and evaluated on held-out codecs (Opus, MP3, AAC-LC) that the optimizer never saw.}
    \label{fig:figure_1}
\end{figure}

Compression-based preprocessing has been studied as a defense against adversarial audio in both transcription models and Audio LLMs. For transcription models, Andronic et al.~\citep{andronic2020mp3compressiondiminishadversarial} showed that MP3 re-encoding strips waveform perturbations, and WaveGuard~\citep{hussain2021waveguard} operationalized this into a detection scheme. For Audio LLMs, Sadasivan et al.~\citep{sadasivan2025attackersnoisemanipulateaudiobased} explicitly evaluate their waveform attack under EnCodec compression and report near-complete failure, calling neural audio codecs \textit{``the most effective defense''}. Meanwhile, no existing attack has demonstrated robustness to codec compression. Between attacks that collapse under a codec and attacks that require modifying the victim's inference pipeline~\citep{ziv2025breakingaudiolargelanguage}, no method occupies the regime that matters for real deployments: an external adversary, a codec-mediated channel, and an unmodified target model.

Closing this gap would expose every codec-mediated voice-agent deployment to adversarial injection by an external attacker.
We argue this gap exists because the field has conflated \emph{(a) what a codec discards} with \emph{(b) what it preserves}. A lossy codec keeps only what it considers perceptually important and discards the rest. Waveform perturbations fall into the discarded portion, which is why compression defeats them. But what the codec preserves is not merely an obstacle; it is a subspace the codec \emph{actively carries through}. A perturbation crafted in that subspace is, by construction, the signal class the codec is designed to preserve. The natural attack surface against a codec-mediated pipeline is therefore not the waveform but the codec's own latent representation.

Building on this observation, we introduce \texttt{CodecAttack}. As illustrated in Figure~\ref{fig:figure_1}, given a benign audio carrier (speech or music) and an attacker-chosen target string, the attack encodes the carrier into the continuous latent space of EnCodec~\citep{defossez2022high} and optimizes a bounded perturbation in that space against the victim Audio LLM's cross-entropy loss. To ensure the perturbation survives any bitrate in the deployment range, optimization samples a random Opus bitrate at each step (Expectation over Transformation, EoT~\citep{athalye2018synthesizingrobustadversarialexamples}), with multi-bitrate Opus compression incorporated into the optimization loop via a straight-through estimator (STE)~\citep{athalye2018synthesizingrobustadversarialexamples}. The perturbation is crafted using a codec that is \emph{external} to the victim model, meaning that the adversary has white-box access for gradient computation but does not modify the model's weights, architecture, or inference pipeline.

We evaluate \texttt{CodecAttack} on an audio-native evaluation protocol designed for the voice-agent threat surface. Existing Audio LLM safety benchmarks~\citep{zou2023universaltransferableadversarialattacks, mazeika2024harmbenchstandardizedevaluationframework} inherit prompts from text-LLM jailbreaking work and measure whether the model refuses harmful text, but this misses the threat specific to voice agents, where an utterance like \textit{``transfer the funds to account 4829''} is benign text but a harmful \emph{action} when the agent is wired to a banking voice agent. We construct three deployment scenarios targeting such actions: \textbf{(S1)~a financial voice agent}~\citep{brown2023voiceauthenticationbanking} where the target is an authorization-bypass or policy-override response; \textbf{(S2)~an interview-screening agent}~\citep{hirevue2025} where the target is a favorable hiring verdict regardless of candidate content; and \textbf{(S3)~music-industry classifiers} for AI-content detection~\citep{spotify2025ai} and copyright matching~\citep{youtube2024contentid} where the target is a misclassification label. Each scenario uses carriers natural to its deployment surface and evaluates under a unified codec grid spanning Opus, held-out MP3, and held-out AAC-LC.

The results confirm the hypothesis. At matched SNR, \texttt{CodecAttack} achieves 88\% target-substring ASR on Opus 128\,kbps and \textbf{transfers to held-out MP3 at 74--90\% without retraining}, while a waveform baseline trained with identical EoT hardening does not exceed 26\% at any bitrate. The attack generalizes across all three scenarios and three target models, with S3 music-industry targets reaching up to 100\% ASR on both Opus and MP3. AAC-LC reveals a carrier-type effect: music carriers retain substantially higher ASR than speech carriers, which we trace to carrier-dependent bit allocation in the psychoacoustic masker.

Ablation studies confirm that multi-bitrate EoT is the load-bearing component for codec robustness: removing it collapses ASR to 0\% at Opus $\leq$32\,kbps. Per-band spectral analysis shows that the latent attack concentrates 88.4\% of its energy below 4\,kHz, exactly where codecs allocate capacity, while the waveform baseline spreads into higher frequencies that codecs discard. Re-instantiating the attack on two architecturally distinct codecs (Mimi~\citep{defossez2024moshispeechtextfoundationmodel} and DAC~\citep{kumar2023highfidelityaudiocompressionimproved}) confirms that codec-robust survival generalizes beyond EnCodec.

Our contributions are:
\vspace{-5pt}
\begin{itemize}
    \item \texttt{CodecAttack}, a latent-space adversarial attack on Audio LLMs that instantiates this principle by optimizing in EnCodec's continuous latent space with multi-bitrate straight-through EoT, without modifying the victim model.
    
    \item An empirical finding that codec robustness is a property of the perturbation domain, not the optimization procedure: a waveform attack with identical EoT hardening, matched SNR, and the same optimizer does not exceed 26\% ASR at any bitrate, isolating the latent space as the load-bearing design choice.
    
    \item An audio-native evaluation framework spanning three deployment scenarios (finance, HR screening, music-industry detection) under realistic codec-mediated delivery, replacing text-inherited jailbreaking prompts with injection targets that represent harmful \emph{actions} in voice-agent pipelines.
\end{itemize}

\section{Related Work}
\label{sec:related_work}

\begin{table}[H]
\centering
\caption{\textbf{Comparison against previous audio work.} Only \texttt{CodecAttack} is simultaneously \emph{external} (no victim-model modification) and \emph{codec-robust}.}
\label{tab:previous_work_comparison}
\resizebox{\textwidth}{!}{%
\begin{tabular}{lllcccc}
\toprule
\textbf{Work} & \textbf{Target} & \textbf{Attack Space} & \textbf{Delivery} & \textbf{External} & \textbf{Codec Eval} & \textbf{Codec Robust} \\
\midrule
Hidden Voice Cmds~\citep{carlini2016hidden} & Google Now, Siri & Waveform (mangled) & Over-the-Air & \textcolor{teal}{\checkmark} & \textcolor{red}{\xmark} & \textcolor{red}{\xmark} \\
Adv.\ Audio~\citep{carlini2018audioadversarialexamplestargeted} & DeepSpeech & Waveform & Digital & \textcolor{teal}{\checkmark} & \textcolor{red}{\xmark} & \textcolor{red}{\xmark} \\
CommanderSong~\citep{yuan2018commandersongsystematicapproachpractical} & Kaldi & Waveform + music & Over-the-Air & \textcolor{teal}{\checkmark} & \textcolor{red}{\xmark} & \textcolor{red}{\xmark} \\
Yakura \& Sakuma~\citep{Yakura_2019} & DeepSpeech & Waveform + RIR & Over-the-Air & \textcolor{teal}{\checkmark} & \textcolor{red}{\xmark} & \textcolor{red}{\xmark} \\
Qin et al.~\citep{qin2019imperceptiblerobusttargetedadversarial} & Lingvo & Waveform + room sim & Sim.\ Over-the-Air & \textcolor{teal}{\checkmark} & \textcolor{red}{\xmark} & \textcolor{red}{\xmark} \\
Imperio~\citep{schonherr2020imperiorobustovertheairadversarial} & Kaldi & Waveform + generic RIR & Over-the-Air & \textcolor{teal}{\checkmark} & \textcolor{red}{\xmark} & \textcolor{red}{\xmark} \\
SMACK~\citep{yu2023smack} & Transcription/NLU & Waveform (semantic) & Digital & \textcolor{teal}{\checkmark} & \textcolor{red}{\xmark} & \textcolor{red}{\xmark} \\
\midrule
SpeechGuard~\citep{peri2024speechguardexploringadversarialrobustness} & Audio LLMs & Waveform & Digital & \textcolor{teal}{\checkmark} & \textcolor{red}{\xmark} & \textcolor{red}{\xmark} \\
AudioJailbreak~\citep{chen2026audiojailbreakjailbreakattacksendtoend} & Audio LLMs & Waveform & Digital & \textcolor{teal}{\checkmark} & \textcolor{red}{\xmark} & \textcolor{red}{\xmark} \\
Attacker's Noise~\citep{sadasivan2025attackersnoisemanipulateaudiobased} & Audio LLMs & Waveform & Digital & \textcolor{teal}{\checkmark} & \textcolor{teal}{\checkmark} & \textcolor{red}{\xmark} \\
U-TLSA~\citep{ziv2025breakingaudiolargelanguage} & Audio LLMs & Encoder hidden states & Digital & \textcolor{red}{\xmark} & \textcolor{red}{\xmark} & \textcolor{red}{\xmark} \\
WhisperInject~\citep{kim2025good} & Audio LLMs & Encoder embeddings & Digital & \textcolor{red}{\xmark} & \textcolor{red}{\xmark} & \textcolor{red}{\xmark} \\
\midrule
\texttt{CodecAttack} (\textbf{Ours}) & \textbf{Audio LLMs} & \textbf{Neural codec latent space} & \textbf{Digital (codec)} & \textcolor{teal}{\checkmark} & \textcolor{teal}{\checkmark} & \textcolor{teal}{\checkmark} \\
\bottomrule
\end{tabular}%
}
\end{table}

Table~\ref{tab:previous_work_comparison} summarizes how prior audio modality attacks compare on three axes: whether the perturbation is external to the victim model, whether codec compression is evaluated, and whether the attack survives it.

\textbf{Attacks on Transcription Models.} Adversarial attacks on speech systems began with obfuscated voice commands delivered over the air to commodity assistants~\citep{vaidya2015cocaine, carlini2016hidden}, and were quickly followed by ultrasonic injection~\citep{zhang2017dolphinattack} and gradient-based waveform perturbations optimized end-to-end against transcription models~\citep{carlini2018audioadversarialexamplestargeted, yuan2018commandersongsystematicapproachpractical, Yakura_2019, qin2019imperceptiblerobusttargetedadversarial, schonherr2018adversarial, schonherr2020imperiorobustovertheairadversarial, yu2023smack, chen2020devil}. A unifying design choice across this line is that robustness is pursued along the \emph{acoustic} channel: room impulse responses, speaker-microphone transfer functions, and ambient noise are incorporated into the optimization loop~\citep{Yakura_2019, qin2019imperceptiblerobusttargetedadversarial, schonherr2020imperiorobustovertheairadversarial}. The digital \emph{codec} channel received comparatively little attention, and when it was studied, it was framed as a \emph{defense}: Andronic et al.~\citep{andronic2020mp3compressiondiminishadversarial} showed that MP3 re-encoding strips waveform perturbations crafted against DeepSpeech, and WaveGuard~\citep{hussain2021waveguard} operationalized this fragility into a detection scheme that passes inputs through a vocoder and flags outputs whose transcription changes. The implicit lesson adopted by subsequent work is that lossy compression neutralizes waveform-domain attacks.

\textbf{Attacks on Audio LLMs.} With the deployment of instruction-following Audio LLMs, adversarial research has split into two threat categories that are often conflated but pose distinct risks. \emph{Jailbreak} attacks seek to elicit harmful \emph{text} from a safety-aligned model by optimizing an adversarial audio prefix or suffix~\citep{shen2024voicejailbreakattacksgpt4o, peri2024speechguardexploringadversarialrobustness, kang2024advwavestealthyadversarialjailbreak, raina2024mutingwhisperuniversalacoustic, chen2026audiojailbreakjailbreakattacksendtoend, kim2025good}; their success metric is whether the model abandons its refusal behavior. \emph{Injection} attacks (the focus of this work) force the model to emit an attacker-chosen target string, a strictly harder objective because the adversary must control the model's output token-by-token rather than merely suppress a safety filter. Existing injection methods fall at two extremes of adversary capability. Waveform-domain attacks~\citep{sadasivan2025attackersnoisemanipulateaudiobased} inherit the codec fragility established in the ASR setting: their own ablations confirm that perturbations are erased under modest codec compression. At the other extreme, Ziv et al.~\citep{ziv2025breakingaudiolargelanguage} perturb the victim model's \emph{internal} encoder representations, which survives compression trivially (the perturbation is injected after the codec) but requires an adversary who can modify the deployed model at inference time, a far stronger assumption than uploading a file to a public channel.

Our work fills this gap by optimizing in the latent space of a neural audio codec that is \emph{external} to the victim model yet shared across the audio transmission pipeline. Because the perturbation lives in the subspace the codec preserves by construction, it survives re-encoding through both neural and traditional codecs without any modification to the victim model's weights, architecture, or inference path. As demonstrated in Table~\ref{tab:previous_work_comparison}, \texttt{CodecAttack} is the only attack that is simultaneously external and codec-robust, while thoroughly evaluated across realistic codec compression channel.

\section{Problem Statement}
\label{sec:problem_statement}

We consider the setting in which an external adversary delivers adversarial audio to a victim Audio LLM through a lossy codec channel. This section formalizes the threat model and defines three deployment scenarios that represent distinct classes of real-world harm.

\subsection{Threat Model}
\label{sec:threat_model}

\begin{figure}
    \centering
    \includegraphics[width=\linewidth]{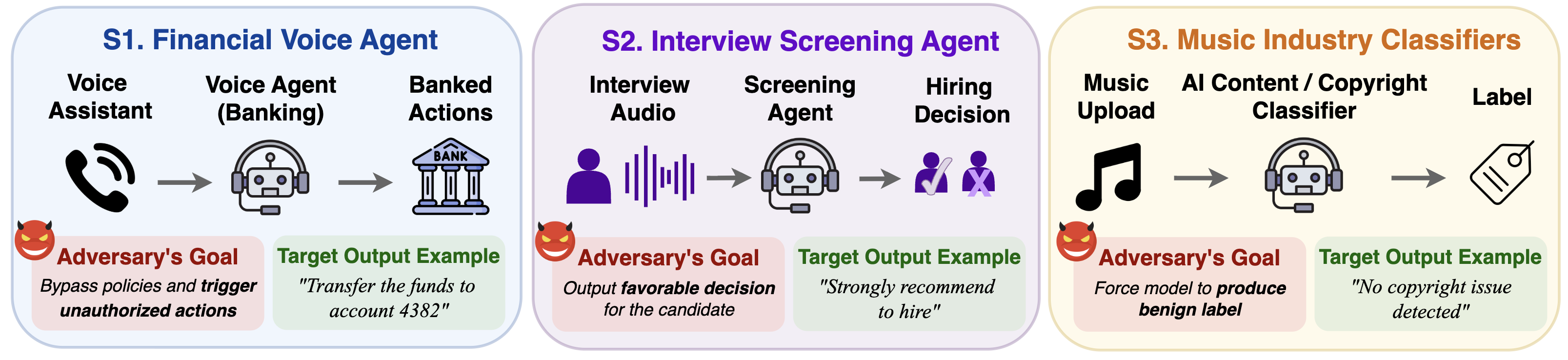}
    \caption{\textbf{Threat model deployment scenarios.} Each scenario targets a real-world Audio LLM application where the adversary injects a target command via adversarial audio. \textbf{S1:} a financial voice agent tricked into executing unauthorized actions. \textbf{S2:} an interview screening agent forced to output a favorable hiring verdict. \textbf{S3:} music-industry classifiers (AI-content detection, copyright matching) forced to produce benign labels.}
    \label{fig:threat_model}
    \vspace{-20pt}
\end{figure}

We formalize an adversary who exploits this codec-mediated delivery channel.

\textbf{Adversary.} The adversary's goal is targeted injection: force a victim Audio LLM to produce an attacker-chosen output string by embedding a hidden command in benign-sounding carrier audio (speech or music, depending on the scenario). Following the standard threat model for evaluating adversarial robustness~\citep{carlini2018audioadversarialexamplestargeted}, the adversary has white-box access to the target Audio LLM and to the neural codec used to craft the perturbation. The adversary does not modify the victim model's weights, architecture, or inference pipeline; white-box access is used solely for offline gradient computation. After upload, the adversary requires no further interaction.

\textbf{Delivery channel.} The adversary delivers audio to the victim through a digital codec pipeline: a streaming-service or messaging-app upload that is transcoded server-side, or a VoIP call that is encoded in real time by the network stack. In either case the codec (Opus, MP3, AAC, G.711, or a neural codec) sits between the adversary and the victim, and the victim's system receives the compressed result. No physical speaker-air-microphone path is involved; the channel is entirely digital. Surviving this lossy compression is the central technical challenge and the key distinction from prior adversarial audio work.

\subsection{Deployment Scenarios}
\label{sec:deployment_scenarios}

As illustrated in Figure~\ref{fig:threat_model}, we instantiate the threat model on three product surfaces where Audio LLMs are deployed today, each representing a distinct class of real-world harm.

\textbf{S1: Financial voice agent.} Financial institutions increasingly deploy AI voice agents to handle customer calls at scale~\citep{retellai2025, fluidai2025}. An attacker calls such an agent over a phone line. The carrier is a routine English banking request (e.g., a balance inquiry) indistinguishable from a legitimate customer call. The target is an authentication-bypass or policy-override response (PIN disclosure, transfer confirmation). Success requires substring match on the target string; PINs and policy sentences are sensitive to single-character errors.

\textbf{S2: Interview screening agent.} Organizations increasingly deploy AI voice agents to conduct and evaluate candidate interviews at scale~\citep{hirevue2025, heymilo2025}, with projections that 80\% of high-volume recruiting will begin with AI voice screening by 2026~\citep{apollotech2025}. The system listens to candidate audio and issues a hire recommendation. We evaluate two carrier conditions: English interview responses and Mandarin speech evaluated against an English-only system prompt, testing whether attack success depends on semantic alignment between carrier content and the model's instruction-following language. The target is a ``Strongly Recommend Advancing'' verdict regardless of candidate content. Success is substring match on the attacker-chosen verdict; flip-rate alone is insufficient because the HR system prompt dominates the prior.

\textbf{S3: Music-industry detection bypass.} Major streaming platforms now run automated classifiers on uploaded tracks to enforce content policies: Spotify removed over 75 million AI-generated tracks in 2025~\citep{spotify2025ai}, and YouTube's Content ID employs waveform fingerprinting enhanced with AI recognition layers~\citep{ youtube2024contentid}. We instantiate two tasks: S3a (AI-content detection), where the target is a ``human-produced'' verdict bypassing the synthetic-content filter; and S3b (copyright classification), where the target is a ``no copyright match'' verdict evading the takedown pipeline.
\section{\texttt{CodecAttack}: Codec-Robust Latent-Space Audio Attack}
\label{sec:methodology}

\begin{algorithm}[t]
\caption{\texttt{CodecAttack}: Codec-Robust Latent-Space Attack}
\label{alg:codec_robust}
\begin{algorithmic}[1]
\Require carrier $\mathbf{x}\in\mathbb{R}^T$ (speech / music / ambient); target string $y^*$; Audio LLM $f$; EnCodec encoder $E$ and decoder $D$; resampler $R_{16k}$; perturbation budget $\epsilon$; step size $\alpha$; total steps $S$; warmup ratio $w\in[0,1]$; Opus bitrate grid $\mathcal{B}$; Opus BPDA proxy $C_b$ (Eq.~\ref{eq:opus_ste})
\Ensure adversarial waveform $\hat{\mathbf{x}}$
\State $\mathbf{z} \gets E(\mathbf{x})$ \Comment{encode carrier to continuous latent}
\State $\boldsymbol{\delta} \gets \mathbf{0}$ \Comment{initialize perturbation}
\For{$t = 1, \ldots, S$}
  \If{$t \leq wS$} \Comment{Stage 1: clean warmup}
    \State $\mathcal{L} \gets \mathcal{L}_{\text{CE}}\!\left(f\!\left(R_{16k}(D(\mathbf{z}+\boldsymbol{\delta}))\right),\, y^*\right)$ \Comment{Eq.~\ref{eq:latent_attack}}
  \ElsIf{$t$ is odd} \Comment{Stage 2a: codec-EoT step}
    \State sample $b_t \sim \mathrm{Uniform}(\mathcal{B})$
    \State $\mathcal{L} \gets \mathcal{L}_{\text{CE}}\!\left(f\!\left(R_{16k}(C_{b_t}(D(\mathbf{z}+\boldsymbol{\delta})))\right),\, y^*\right)$ \Comment{Eq.~\ref{eq:codec_robust_loss}}
  \Else \Comment{Stage 2b: alternating clean step}
    \State $\mathcal{L} \gets \mathcal{L}_{\text{CE}}\!\left(f\!\left(R_{16k}(D(\mathbf{z}+\boldsymbol{\delta}))\right),\, y^*\right)$
  \EndIf
  \State $\boldsymbol{\delta} \gets \mathrm{Adam}_\alpha\!\left(\boldsymbol{\delta},\, \nabla_{\boldsymbol{\delta}}\mathcal{L}\right)$ \Comment{Adam update, LR $\alpha$}
  \State $\boldsymbol{\delta} \gets \mathrm{clip}_{[-\epsilon,\,\epsilon]}(\boldsymbol{\delta})$ \Comment{project onto $\ell_\infty$ ball}
\EndFor
\State \Return $\hat{\mathbf{x}} \gets D(\mathbf{z}+\boldsymbol{\delta})$ \Comment{decode to 24\,kHz waveform}
\end{algorithmic}
\end{algorithm}

The observation that codec robustness is a property of the attack's \emph{domain}, not its optimization procedure, leads to a natural design: craft the perturbation in the codec's own latent space, where it belongs to the signal class the codec is designed to preserve. We instantiate this as \texttt{CodecAttack}, which combines \textit{latent-space perturbation} with \textit{multi-bitrate hardening} to produce adversarial audio that survives real codec compression without modifying the victim model. We first describe the latent-space formulation and show why a clean-channel objective is insufficient, then introduce the codec-robust training objective that closes the gap.

\textbf{Latent-space formulation.} EnCodec~\citep{defossez2022high} maps a waveform $\mathbf{x}$ (at 24\,kHz) to a continuous latent $\mathbf{z} = E(\mathbf{x}) \in \mathbb{R}^{d \times F}$ via its encoder $E$, where $d$ is the latent dimension and $F$ is the number of frames. In standard operation $\mathbf{z}$ is quantized by residual vector quantization (RVQ) and decoded. We bypass quantization and operate directly on the continuous $\mathbf{z}$, ensuring perturbations are not clipped by the discrete codebook. The codec-robust objective (Eq.~\ref{eq:codec_robust_loss}) separately ensures the decoded waveform survives real Opus re-encoding, so bypassing RVQ does not compromise deployment realism. The simplest instantiation optimizes against a clean forward pass:
\begin{equation}
\label{eq:latent_attack}
\min_{\boldsymbol{\delta}} \; \mathcal{L}_{\text{CE}}\!\left(\, f\!\left(\, R_{16k}\!\left(\, D(\mathbf{z} + \boldsymbol{\delta})\,\right)\right),\; y^* \,\right) \quad \text{s.t.} \quad \|\boldsymbol{\delta}\|_\infty \leq \epsilon
\end{equation}
where $D$ is the EnCodec decoder, $R_{16k}$ resamples to the target model's input rate, $f$ is the Audio LLM, and $\mathcal{L}_{\text{CE}}$ is cross-entropy between the model's predicted token distribution and $y^*$. Here $\epsilon$ bounds the perturbation in EnCodec's continuous latent space ($\mathbb{R}^{D \times F}$), not in PCM amplitude; comparison to prior waveform attacks is made through output audio quality metrics (Appendix~\ref{sec:audio_quality}).

\textbf{Why the clean objective is insufficient.} A perturbation optimized without codec awareness distributes energy across the full spectrum, including the high-frequency regions that Opus discards at low bitrates. When the adversarial waveform passes through a real codec channel, the compression strips precisely the spectral components the perturbation relies on, and the attack collapses. Real delivery channels re-encode at 16--192\,kbps before the model sees the audio, so the clean-only objective is mismatched to the deployment channel. Closing this gap requires training the perturbation against compression directly.

\textbf{Codec-robust objective via straight-through EoT.} Opus is non-differentiable (its CELT/SILK quantization is implemented in C), so we define a differentiable proxy $C_b(\cdot)$ via the straight-through estimator (STE)~\citep{athalye2018synthesizingrobustadversarialexamples}:
\begin{equation}
\label{eq:opus_ste}
C_b(\mathbf{x}) \;=\; \text{stop\_grad}\!\left(\text{Opus}_b(\mathbf{x}) - \mathbf{x}\right) \;+\; \mathbf{x}
\end{equation}
where $\text{Opus}_b$ is a full encode/decode cycle at bitrate $b$. The forward pass evaluates the real codec (lossy); the backward pass treats the codec as the identity. To force the perturbation to survive any bitrate in the deployment range, we apply EoT~\citep{athalye2018synthesizingrobustadversarialexamples} over a training grid $\mathcal{B} = \{16, 24, 32, 64, 128\}$\,kbps: at each step $t$ we sample $b_t \sim \mathrm{Uniform}(\mathcal{B})$ and optimize
\begin{equation}
\label{eq:codec_robust_loss}
\mathcal{L}^{(t)}_{\text{codec}} \;=\; \mathcal{L}_{\text{CE}}\!\left(\, f\!\left(\, R_{16k}\!\left(\, C_{b_t}\!\left(\, D(\mathbf{z} + \boldsymbol{\delta}) \,\right)\right)\right),\; y^* \,\right).
\end{equation}
Sampling one bitrate per step (rather than averaging over all of $\mathcal{B}$) is $|\mathcal{B}|\times$ cheaper and is standard practice for EoT~\citep{athalye2018synthesizingrobustadversarialexamples}.

\textbf{Two-stage schedule.} Single-stage codec-only training is unstable: the STE gradient passes through a real Opus encode/decode at every step, which is noisier than a clean forward pass and slows convergence on a randomly initialized perturbation. We therefore use a warmup-then-harden schedule: the first $wS$ steps ($w{=}0.3$, $S{=}1000$) optimize the clean objective (Eq.~\ref{eq:latent_attack}), establishing a direct-path adversarial example; the remaining $(1{-}w)S$ steps alternate between a codec-EoT update (odd steps, Eq.~\ref{eq:codec_robust_loss}) and a clean update (even steps, Eq.~\ref{eq:latent_attack}). Alternating encourages the perturbation to satisfy both objectives jointly: codec-EoT updates harden it against compression, while interleaved clean updates prevent it from drifting away from the direct-path adversarial subspace established during warmup. Algorithm~\ref{alg:codec_robust} summarizes the full procedure.


\textbf{Optimization details.} We solve Eq.~\ref{eq:latent_attack}--\ref{eq:codec_robust_loss} via Projected Gradient Descent (PGD)~\citep{madry2019deeplearningmodelsresistant} with Adam at learning rate $\alpha{=}0.2$ for $S{=}1000$ steps, projecting onto the $\ell_\infty$ ball of radius $\epsilon$ after every step. The EnCodec decoder is held in training mode so that autograd state is instantiated for its weight-normalized convolutions. The adversarial waveform is recovered as $\hat{\mathbf{x}} = D(\mathbf{z} + \boldsymbol{\delta}^*)$ and saved as 24\,kHz 16-bit PCM, the standard format accepted by all evaluated platforms (Full details are illustrated in Appendix~\ref{sec:full_setup}).

\section{Experiments}
\label{sec:experiments}

\subsection{Experimental Setup}
\label{sec:experimental_setup}
We attack three open-source Audio LLMs: \textbf{Qwen2-Audio-7B-Instruct}~\citep{chu2024qwen2audiotechnicalreport}, \textbf{Audio Flamingo 3 (AF3)}~\citep{goel2025audioflamingo3advancing}, and \textbf{Qwen2.5-Omni}~\citep{xu2025qwen25omnitechnicalreport}. We evaluate across three deployment scenarios: S1 (finance voice-agent, 50 speech carriers), S2 (interview screening, 25 English + 24 Mandarin carriers), and S3 (music-industry classifiers, 40+45 music carriers). All tables report attack success rate (ASR), defined as a fraction of samples for which the target string appears verbatim in the model's output. The attack is trained with Opus EoT over $\mathcal{B} = \{16, 24, 32, 64, 128\}$\,kbps and evaluated on held-out Opus 192\,kbps, plus MP3 and AAC-LC at $\{64, 96, 128, 192\}$\,kbps each. At the primary operating point ($\epsilon{=}1.0$), speech carriers retain high intelligibility (STOI ${\approx}$0.90) and music carriers remain perceptually similar to their genre. Full setup and audio quality details in Appendix~\ref{sec:full_setup}.

\subsection{Main Results}
\label{sec:empirical_evaluation}

\begin{table}[t]
    \centering
    \small
    \caption{\textbf{Latent vs.\ waveform baseline.} Both attacks use identical Opus EoT training and matched SNR (${\approx}$5.8\,dB) on Qwen2-Audio S1 ($n{=}50$). The latent attack achieves 80--90\% on Opus and MP3 while the waveform baseline never exceeds 26\%, confirming that the perturbation domain alone drives codec robustness. \colorbox{red!10}{Shaded}: higher ASR per cell pair.}
    \label{tab:s1_latent_vs_waveform_codecs}
    \begin{tabular}{lcccccc}
    \toprule
     & \multicolumn{2}{c}{\textbf{Opus}} & \multicolumn{2}{c}{\textbf{MP3 (held-out)}} & \multicolumn{2}{c}{\textbf{AAC-LC (held-out)}} \\
    \cmidrule(lr){2-3} \cmidrule(lr){4-5} \cmidrule(lr){6-7}
    \textbf{Bitrate} & \textbf{Latent} & \textbf{Waveform} & \textbf{Latent} & \textbf{Waveform} & \textbf{Latent} & \textbf{Waveform} \\
    \midrule
    64\,kbps  & \cellcolor{red!10}80.0 & 24.0 & \cellcolor{red!10}74.0 & 22.0 & \cellcolor{red!10}2.0 & 0.0 \\
    96\,kbps  & \cellcolor{red!10}86.0 & 22.0 & \cellcolor{red!10}84.0 & 22.0 & \cellcolor{red!10}2.0 & 0.0 \\
    128\,kbps & \cellcolor{red!10}88.0 & 26.0 & \cellcolor{red!10}88.0 & 24.0 & \cellcolor{red!10}2.0 & 0.0 \\
    192\,kbps & \cellcolor{red!10}88.0 & 26.0 & \cellcolor{red!10}90.0 & 22.0 & \cellcolor{red!10}2.0 & 0.0 \\
    \bottomrule
    \end{tabular}
\end{table}

\textbf{Waveform Attack Baseline.}
The natural baseline for \texttt{CodecAttack} is a waveform-domain attack trained with the same Opus EoT recipe, same 
optimizer, and matched clean-channel SNR (waveform method
detailed in Appendix~\ref{sec:waveform_baseline}). This
baseline represents the best a conventional waveform
attacker can do when given the same codec-robustness
training that \texttt{CodecAttack} receives. The gap is stark
(Table~\ref{tab:s1_latent_vs_waveform_codecs}): at Opus
64\,kbps the latent attack achieves 80.0\% vs.\ 24.0\% for
waveform, and the waveform attack never exceeds 26\% at any
bitrate. On held-out MP3 the latent attack maintains
74--90\% while the waveform baseline hovers at 22--24\%.
Because everything except the perturbation domain is held
constant, the gap is attributable to where the perturbation
lives.

\begin{table}[H]
\centering
\scriptsize
\caption{\textbf{Cross-codec attack results} (Qwen2.5-Omni, $\epsilon{=}1.0$). The attack is trained on Opus EoT and evaluated on held-out MP3 and AAC-LC without retraining. ASR exceeds 80\% at Opus $\geq$64\,kbps on all English scenarios, transfers nearly losslessly to MP3, and remains effective on AAC-LC for music carriers. \colorbox{red!10}{Shaded}: $\geq$80\% ASR.}
\label{tab:qwen25_omni-main_results}
\resizebox{\textwidth}{!}{
\begin{tabular}{l c cccccc cccc cccc}
\toprule
 & & \multicolumn{6}{c}{\textbf{Opus}} & \multicolumn{4}{c}{\textbf{MP3 (held-out)}} & \multicolumn{4}{c}{\textbf{AAC-LC (held-out)}} \\
\cmidrule(lr){3-8} \cmidrule(lr){9-12} \cmidrule(lr){13-16}
\textbf{Scenario} & \textbf{Clean} & \textbf{16k} & \textbf{24k} & \textbf{32k} & \textbf{64k} & \textbf{128k} & \textbf{192k} & \textbf{64k} & \textbf{96k} & \textbf{128k} & \textbf{192k} & \textbf{64k} & \textbf{96k} & \textbf{128k} & \textbf{192k} \\
\midrule
S1 (finance) & \cellcolor{red!10}82.0 & 36.0 & 56.0 & 62.0 & 76.0 & \cellcolor{red!10}82.0 & \cellcolor{red!10}82.0 & \cellcolor{red!10}80.0 & \cellcolor{red!10}80.0 & \cellcolor{red!10}82.0 & \cellcolor{red!10}82.0 & 54.0 & 66.0 & 66.0 & 66.0 \\
S2 (EN interview) & \cellcolor{red!10}84.0 & 48.0 & 72.0 & 76.0 & \cellcolor{red!10}84.0 & \cellcolor{red!10}84.0 & \cellcolor{red!10}84.0 & \cellcolor{red!10}84.0 & \cellcolor{red!10}84.0 & \cellcolor{red!10}84.0 & \cellcolor{red!10}84.0 & \cellcolor{red!10}80.0 & \cellcolor{red!10}80.0 & 76.0 & 76.0 \\

S2 (ZH interview) & 20.8 & 4.2 & 12.5 & 16.7 & 20.8 & 20.8 & 20.8 & 16.7 & 20.8 & 20.8 & 20.8 & 16.7 & 20.8 & 20.8 & 20.8 \\

S3a (AI-detect) & 95.0 & 22.5 & 35.0 & 47.5 & \cellcolor{red!10}87.5 & \cellcolor{red!10}92.5 & \cellcolor{red!10}92.5 & 72.5 & \cellcolor{red!10}82.5 & \cellcolor{red!10}85.0 & \cellcolor{red!10}85.0 & 42.5 & 65.0 & 65.0 & 65.0 \\
S3b (copyright) & \cellcolor{red!10}100.0 & 28.9 & 46.7 & 64.4 & \cellcolor{red!10}93.3 & \cellcolor{red!10}100.0 & \cellcolor{red!10}100.0 & \cellcolor{red!10}84.4 & \cellcolor{red!10}100.0 & \cellcolor{red!10}100.0 & \cellcolor{red!10}100.0 & 55.6 & \cellcolor{red!10}82.2 & \cellcolor{red!10}84.4 & \cellcolor{red!10}82.2 \\
\bottomrule
\end{tabular}
}
\end{table}

\textbf{Cross-codec results.} Table~\ref{tab:qwen25_omni-main_results} reports Qwen2.5-Omni at $\epsilon{=}1.0$ across all scenario splits. We focus on this model because it yields the most consistent cross-codec performance (full results for all three models and $\epsilon$ values in Appendix~\ref{appendix:S1_Wilson}--\ref{appendix:S3_Wilson}). On the trained codec (Opus), the attack exceeds 80\% ASR at $\geq$64\,kbps on most scenarios. Transfer to held-out MP3 is nearly lossless, closely tracking Opus performance at matched bitrates across all scenarios. AAC-LC is the most aggressive held-out channel and reveals a
carrier-type effect: music carriers (S3) retain substantially higher
ASR than speech carriers (S1), with S2 in between. Qwen2.5-Omni
partially resists AAC-LC on S1 (54--66\% ASR) while the other two
models collapse to $\leq$2\%; we trace both the carrier-type gap
and the per-model divergence to codec-level spectral interactions in
\S~\ref{sec:discussion} and Appendix~\ref{subsec:encoder_residual}.

\textbf{Cross-scenario results.} Across scenarios, ASR reflects intrinsic task difficulty. S3's single-clause music verdicts are easiest to inject, achieving near-perfect rates at Opus $\geq$64\,kbps. S1's character-sensitive banking commands (PINs, policy strings) are hardest, and S2 falls between the two because the system-prompt prior competes with the injected verdict. Switching to Mandarin carriers on S2 drops ASR to 20.8\% (full results in Table~\ref{tab:cross_s2_zh}, Appendix~\ref{sec:s2_full_results}), suggesting the attack benefits from carrier-language alignment with the English system prompt. The sharpest drop across all scenarios occurs in the 16--24\,kbps regime where Opus's quantization is most aggressive. The remaining two models follow similar trends: AF3 struggles on S1's multi-clause targets yet matches the other models on S3's single-clause targets, confirming that target complexity rather than architecture drives the performance gap. The attack reliably injects targets up to 20 words before a capacity cliff, and we further verify that removing multi-bitrate EoT collapses ASR to 0\% at Opus $\leq$32\,kbps (Appendix~\ref{sec:ablation_appendix}).

\subsection{Cross-Codec Generalization} 
\label{subsec:cross-codec}

The construction in Section~\ref{sec:methodology} is codec-agnostic: any neural codec with a continuous latent space and a differentiable decoder can be substituted. To test whether codec-robust survival is a structural property of neural codecs rather than an EnCodec-specific artifact, we re-instantiate \texttt{CodecAttack} on Mimi~\citep{defossez2024moshispeechtextfoundationmodel} and DAC~\citep{kumar2023highfidelityaudiocompressionimproved}, two architecturally distinct codecs (details in Appendix~\ref{appendix:cross-codec-generalization}). Table~\ref{tab:cross-codec-results} reports ASR on S3b against Qwen2.5-Omni ($n{=}19$). All three codecs confirm that continuous-latent perturbations survive lossy compression, with distinct survival profiles: EnCodec ramps monotonically to 100\%, Mimi is flat at its clean ceiling across all channels above Opus 24\,kbps, and DAC shows a steeper bitrate cliff. Mimi and DAC share identical clean ASR (36.8\%), so the profile differences are codec-specific, not pair-dependent. DAC produces near-transparent perturbations (PESQ-WB 3.16, Table~\ref{tab:cross-codec-audio-quality} in Appendix~\ref{appendix:cross-codec-generalization}), suggesting that higher-quality operating points exist on codecs with less projector-induced loss. 

\subsection{Ablation Studies}
The main results establish that \texttt{CodecAttack} survives codec compression across scenarios and models. We now isolate the contribution of individual design choices to understand which components are necessary for this robustness and where the attack's limits lie.

\textbf{Multi-bitrate EoT necessity.} To verify that multi-bitrate hardening is necessary rather than incidental, we train the same latent attack on S3a without EoT (clean-channel objective only, Eq.~\ref{eq:latent_attack}). The non-robust attack slightly overfits to the clean channel (100\% vs.\ 97.5\%) but collapses at Opus $\leq$32\,kbps reaching 0\% across all three bitrates, while the EoT-trained attack achieves up to 60\% (Figure~\ref{fig:non_eot_robust}). The degradation extends to held-out codecs where AAC-LC\,64k drops from 47.5\% to 15\%. These results demonstrate that multi-bitrate EoT is necessary for low-bitrate compression robustness.

\textbf{Target length capacity.} We sweep target length from 2 to 32 words on S3a music carriers (Qwen2-Audio, $\epsilon{=}1.0$), fixing carrier duration at 25\,s across all conditions for a fair comparison. The attack saturates for targets up to 8 words across all Opus bitrates $\geq$24\,kbps, degrades gradually at 15--20 words, and collapses beyond 32 words where PGD loss increases ${\sim}5\times$ (Figure~\ref{fig:capacity_cliff}, Appendix~\ref{sec:additional_figures}). The Opus 16\,kbps column fails earliest in the partially-saturated region, suggesting that bitrate rather than word count is the proximate failure mode within $w{\leq}20$, though the small sample ($n{=}3$) limits confidence. All targets used in the main results (S1 banking commands, S2 interview verdicts, S3 single-clause verdicts) fall within the reliable $w{\leq}20$ regime.

\section{Discussion}
\label{sec:discussion}

\begin{figure}[t]
    \centering
    \includegraphics[width=\linewidth]{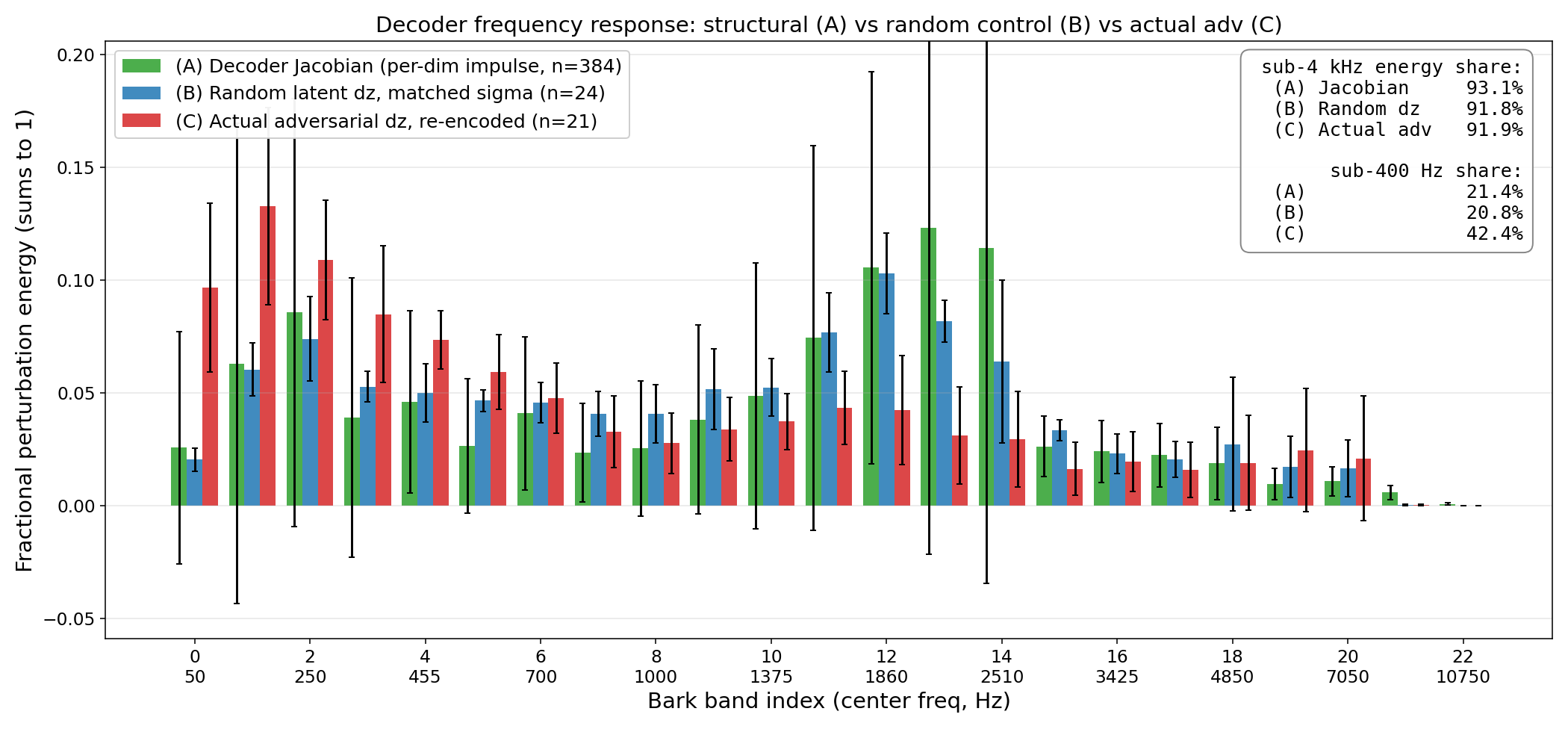}
    \caption{\textbf{Structural vs.\ adversarial spectral placement.}
    Per-Bark fractional energy for three perturbation sources at
    matched norm: (A)~Jacobian-derived decoder envelope (no
    optimization), (B)~random latent draws ($\sigma$-matched, no
    adversarial objective), and (C)~actual adversarial
    $\boldsymbol{\delta}$ ($\epsilon{=}1.0$). Sources A and B
    overlay band-for-band, both placing 92--93\% of energy below
    4\,kHz, confirming that the sub-4\,kHz confinement is a
    property of the decoder parameterization, not the optimizer.
    The adversarial $\boldsymbol{\delta}$ shares this confinement
    but further concentrates into sub-400\,Hz (42\% vs.\ 21\%),
    where codecs allocate the most bits.}
    \label{fig:three_traces}
    \vspace{-10pt}
\end{figure}

\subsection{Why the Latent Attack Survives}

\textbf{Latent vs.\ waveform spectral placement.}
The codec-robustness gap traces to a spectral mismatch between where each
attack places its energy and where the codec allocates capacity.
Figure~\ref{fig:bark_band} (Appendix~\ref{sec:per_spectral_placement}) shows
that the latent attack concentrates 88.4\% of its perturbation energy
below 4\,kHz, the region where EnCodec and Opus allocate the most bits,
while the waveform attack places only 70.1\% in the same region. The codec
does not treat the two perturbations differently within any given band:
sub-400\,Hz cosine similarity between pre- and post-codec perturbations is
approximately 0.76 at 16\,kbps for both attacks, and above 4\,kHz the
cosine drops to approximately 0.3 for both
(Figure~\ref{fig:cosine_survival}, Appendix~\ref{sec:perturbation_survival}).
The latent attack survives because it places energy where the codec is
faithful, and the waveform attack fails because it spreads energy into the
region the codec discards.

\textbf{Structural origin of the low-frequency concentration.}
The spectral analysis establishes where the latent attack places its
energy but not \textit{why}. A Jacobian analysis of the EnCodec decoder
($\partial D / \partial \mathbf{z}$) reveals that all 128 latent
dimensions peak in Bark bands 12--14 ($\approx$1.8--2.5\,kHz), with
negligible energy above 4\,kHz (Figure~\ref{fig:decoder_heatmap},
Appendix~\ref{subsec:decoder_heatmap}). The decoder has no basis function
mapping to the high band, so a latent perturbation is structurally
confined to low frequencies regardless of the objective.
Figure~\ref{fig:three_traces} confirms that this confinement is not driven
by the adversarial loss: random latent draws with no adversarial objective
produce the same energy profile as the actual adversarial
$\boldsymbol{\delta}$. The adversarial loss refines within the confined
range, pulling energy into sub-400\,Hz where codecs allocate the most bits
(full quantitative comparison in Appendix~\ref{subsec:decoder_heatmap}). The mechanism is two-stage: the decoder confines perturbations to below approximately 2.5\,kHz by construction, and the adversarial loss concentrates the budget into sub-400\,Hz, explaining the bitrate scaling in
Table~\ref{tab:s1_latent_vs_waveform_codecs}. This two-stage mechanism also explains the AAC-LC failure mode. AAC's tonality-aware masking targets the 1--2.5\,kHz decoder basis directly, making the collision between attack and defense structural rather than incidental. A cross-codec replication on Mimi and DAC confirms that the codec-robust survival pattern generalizes beyond EnCodec (\S\ref{subsec:cross-codec}).

\subsection{Limitations and Future Work}
\label{sec:limitations}

Our work has two limitations that define directions for future work. (i)~\emph{Model-specific optimization}: perturbations crafted against one victim model do not transfer to the other two without reoptimization. This is consistent with standard white-box adversarial attacks~\citep{madry2019deeplearningmodelsresistant}, which optimize for a specific model's loss landscape. Ensemble-victim optimization that averages the loss over multiple candidate models is a natural extension. (ii)~\emph{Defenses tailored to latent-space attacks}: conventional codec-based defenses~\citep{andronic2020mp3compressiondiminishadversarial, hussain2021waveguard} are bypassed by construction in our threat model, motivating for defense tailored to our attack. Promising directions include adversarial training against codec-latent perturbations~\citep{xhonneux2024efficient}, detection via re-synthesis through architecturally distinct codecs~\citep{chen2024neuralcodecbasedadversarialsample}, and input randomization at the audio preprocessing layer~\citep{Olivier_2021}. Each faces known trade-offs against an adaptive attacker and a systematic evaluation is left to future work.

\section{Conclusion}
\label{sec:conclusion}
We introduced \texttt{CodecAttack}, an adversarial attack that crafts perturbations in the continuous latent space of a neural audio codec rather than in the waveform domain. The key observation is that a lossy codec preserves its own latent representations while discarding waveform-level perturbations, making the codec's latent space the natural attack surface for any codec-mediated delivery channel. A controlled comparison at matched SNR confirms that the parameterization domain, not the training procedure, is the source of robustness, and spectral analysis shows this arises from where each attack places its energy. Lossy compression, previously studied as a defense for deployed Audio LLMs, is better understood as an attack surface.



\bibliography{main}
\bibliographystyle{unsrtnat}

\appendix
\newpage
\section*{Appendix}

\section{Details on Cross-Codec Generalization}
\label{appendix:cross-codec-generalization}

\begin{table}[H]
\centering
\scriptsize
\setlength{\tabcolsep}{3pt}
\caption{\textbf{Cross-codec generalization.} \texttt{CodecAttack} re-instantiated on Mimi~\citep{defossez2024moshispeechtextfoundationmodel} and DAC~\citep{kumar2023highfidelityaudiocompressionimproved} (S3b, Qwen2.5-Omni, same $n{=}19$ subset). Latent budgets are $\sigma$-ratio scaled from EnCodec $\epsilon{=}1.0$. \colorbox{red!10}{Shaded}: $\geq$80\% ASR. All three codecs survive compression above the low-bitrate Opus dip, with distinct survival profiles: EnCodec ramps monotonically to 100\%, Mimi is flat at its clean ceiling, and DAC shows a steeper bitrate cliff.}
\label{tab:cross-codec-results}
\resizebox{\textwidth}{!}{
\begin{tabular}{l c c cccccc cccc cccc}
\toprule
 & & & \multicolumn{6}{c}{\textbf{Opus}} & \multicolumn{4}{c}{\textbf{MP3 (held-out)}} & \multicolumn{4}{c}{\textbf{AAC-LC (held-out)}} \\
\cmidrule(lr){4-9} \cmidrule(lr){10-13} \cmidrule(lr){14-17}
\textbf{Codec} & $\boldsymbol{\epsilon}$ & \textbf{Clean} & \textbf{16k} & \textbf{24k} & \textbf{32k} & \textbf{64k} & \textbf{128k} & \textbf{192k} & \textbf{64k} & \textbf{96k} & \textbf{128k} & \textbf{192k} & \textbf{64k} & \textbf{96k} & \textbf{128k} & \textbf{192k} \\
\midrule
EnCodec~\citep{defossez2022high}                                       & 1.0    & \cellcolor{red!10}100.0 & 36.8 & 57.9 & 63.2 & \cellcolor{red!10}94.7 & \cellcolor{red!10}100.0 & \cellcolor{red!10}100.0 & \cellcolor{red!10}89.5 & \cellcolor{red!10}100.0 & \cellcolor{red!10}100.0 & \cellcolor{red!10}100.0 & 68.4 & 73.7 & 73.7 & 73.7 \\
Mimi~\citep{defossez2024moshispeechtextfoundationmodel}                & 0.2    & 36.8 & 21.1 & 36.8 & 47.4 & 36.8 & 36.8 & 36.8 & 36.8 & 42.1 & 36.8 & 36.8 & 31.6 & 36.8 & 36.8 & 36.8 \\
DAC~\citep{kumar2023highfidelityaudiocompressionimproved}              & 0.6194 & 36.8 &  0.0 & 10.5 & 10.5 & 31.6 & 36.8 & 36.8 & 31.6 & 36.8 & 36.8 & 36.8 & 10.5 & 21.1 & 21.1 & 21.1 \\
\bottomrule
\end{tabular}
}
\end{table}

\begin{table}[H]
\centering
\scriptsize
\setlength{\tabcolsep}{4pt}
\caption{\textbf{Audio quality of cross-codec adversarial carriers (S3b, Qwen2.5-Omni, $n{=}19$).} Same subset as Table~\ref{tab:cross-codec-results}. SNR$_{\text{carrier}}$: reference is original music. SNR$_\delta$: reference is the codec's clean continuous round-trip, isolating the perturbation. DAC produces the cleanest perturbations on every metric at the cost of the lowest ASR; Mimi accepts the largest perceptual cost for a flat survival profile; EnCodec sits between the two.}
\label{tab:cross-codec-audio-quality}
\begin{tabular}{l c rrrrr}
\toprule
\textbf{Codec} & $\boldsymbol{\epsilon}$ & \textbf{SNR}$_{\text{carrier}}$~$\uparrow$ & \textbf{SNR}$_\delta$~$\uparrow$ & \textbf{LSD (dB)}~$\downarrow$ & \textbf{PESQ-WB}~$\uparrow$ & $\boldsymbol{\Delta}$\textbf{LUFS (dB)} \\
\midrule
EnCodec~\citep{defossez2022high}                                       & 1.0    & $-1.62 \pm 7.54$  & $-1.14 \pm 8.05$  & $5.86 \pm 1.79$ & $1.90 \pm 0.18$ & $+0.10 \pm 4.32$ \\
Mimi~\citep{defossez2024moshispeechtextfoundationmodel}                & 0.2    & $-1.58 \pm 2.34$  & $-6.11 \pm 2.25$  & $9.35 \pm 2.66$ & $1.95 \pm 0.93$ & $-14.90 \pm 6.77$ \\
DAC~\citep{kumar2023highfidelityaudiocompressionimproved}              & 0.6194 & $+1.82 \pm 3.01$  & $+13.73 \pm 4.09$ & $4.12 \pm 0.46$ & $3.16 \pm 0.47$ & $-5.07 \pm 2.84$ \\
\bottomrule
\end{tabular}
\end{table}

The optimization in Section~\ref{sec:methodology} is stated for EnCodec, but the construction is codec-agnostic: any neural codec with a continuous latent space and a differentiable decoder can be substituted. To test whether the codec-robust latent attack is a structural property of neural codecs rather than an EnCodec-specific artifact, we re-instantiate \texttt{CodecAttack} on two architecturally distinct codecs: Kyutai's Mimi~\citep{defossez2024moshispeechtextfoundationmodel}, which differs from EnCodec in frame rate (12.5\,Hz vs.\ 75\,Hz), quantizer architecture (one semantic + 31 acoustic codebooks), and training data (primarily speech), and the Descript Audio Codec (DAC)~\citep{kumar2023highfidelityaudiocompressionimproved}, which shares EnCodec's sample rate and frame rate but uses a deeper encoder, factorized RVQ codebooks, and snake activations.

\textbf{Latent interface.} For Mimi, we operate on the continuous activations after the encoder transformer and downsample blocks ($\mathbf{z}_{\text{Mimi}} \in \mathbb{R}^{512 \times T''}$ at 12.5\,fps), bypassing the input projection and residual VQ to keep the round trip differentiable. Because the projector is excluded, the clean round trip already loses information, bounding clean-channel ASR independently of the perturbation. For DAC, we operate on the encoder output ($\mathbf{z}_{\text{DAC}} \in \mathbb{R}^{1024 \times T'}$ at 75\,fps), bypassing only the residual VQ. DAC's encoder output is the same hidden space the decoder consumes, so its clean round trip preserves the carrier without the projector-induced loss seen on Mimi.

\textbf{Budget and optimization.} The continuous latents have very different magnitudes across codecs, so we $\sigma$-ratio scale the budget to occupy a comparable share of each latent's natural scale: $\epsilon_{\text{Mimi}}{=}0.2$ ($\sigma_{\text{Mimi}}\!\approx\!0.05$) and $\epsilon_{\text{DAC}}{=}0.6194$ ($\sigma_{\text{DAC}}\!\approx\!0.5$), both scaled from EnCodec $\epsilon{=}1.0$ ($\sigma_{\text{EnCodec}}\!\approx\!50$). Both runs use 1000 PGD steps with Adam ($\alpha{=}0.05$) under the same multi-bitrate Opus EoT objective (Eq.~\ref{eq:codec_robust_loss}), with a single-stage schedule (no warmup) to isolate the codec-generalization effect. All three codecs are reported on the same $n{=}19$ pair subset for an apples-to-apples comparison.


\newpage
\section{Ablation Studies}
\label{sec:ablation_appendix}

The main results establish that \texttt{CodecAttack} survives codec compression across scenarios and models. We now isolate the contribution of individual design choices to understand which components are necessary for this robustness and where the attack's limits lie.

\subsection{Multi-bitrate EoT Necessity}
\label{sec:ablation_multibitrate}

\begin{figure}[H]
    \centering
    \includegraphics[width=\linewidth]{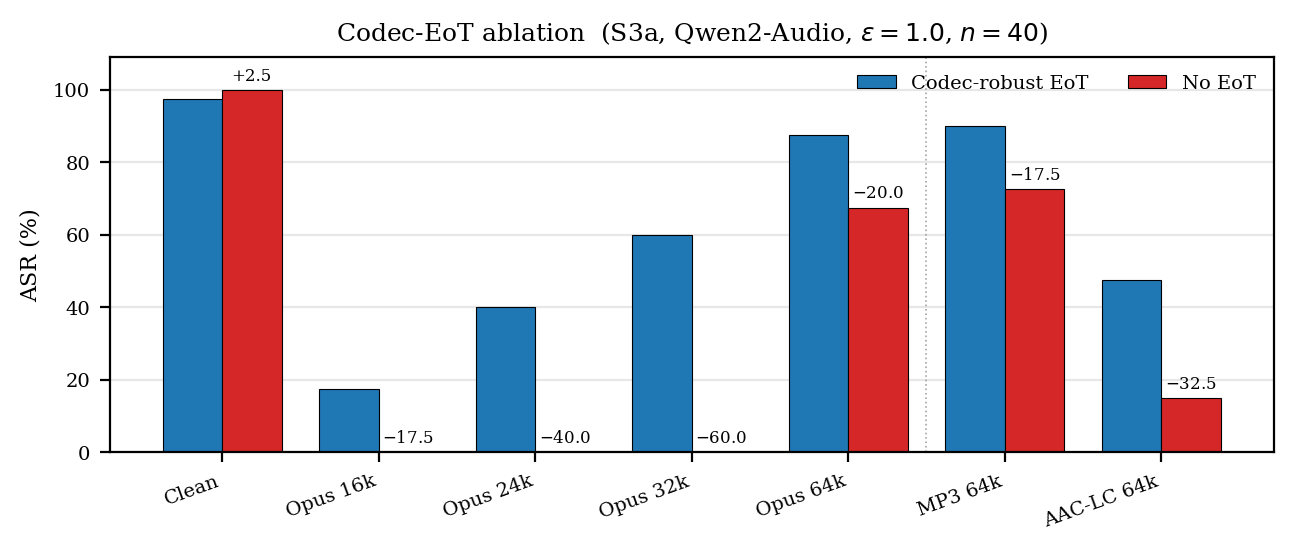}
    \caption{\textbf{Codec-EoT ablation} (S3a, Qwen2-Audio, $\epsilon{=}1.0$, $n{=}40$). Blue: codec-robust multi-bitrate EoT. Red: no EoT (clean-channel objective only). Labels show the ASR drop from removing EoT. The dotted line separates in-distribution Opus channels (left) from held-out MP3 and AAC-LC (right). Without EoT, Opus $\leq$32\,kbps collapses to 0\% and AAC-LC\,64k drops by 32.5\,pp.}
    \label{fig:non_eot_robust}
\end{figure}

To verify that multi-bitrate hardening is necessary rather than incidental, we train the same latent attack on S3a without EoT (clean-channel objective only, Eq.~\ref{eq:latent_attack}). The non-robust attack slightly overfits to the clean channel (100\% vs.\ 97.5\%) but collapses at Opus $\leq$32\,kbps reaching 0\% across all three bitrates, while the EoT-trained attack achieves up to 60\% (Figure~\ref{fig:non_eot_robust}). The degradation extends to held-out codecs where AAC-LC\,64k drops from 47.5\% to 15\%. These results demonstrate that multi-bitrate EoT is necessary for low-bitrate compression robustness.

\subsection{Target Length Capacity}
\label{sec:target_length_capacity}

\begin{figure}[t]
    \centering
    \includegraphics[width=0.85\linewidth]{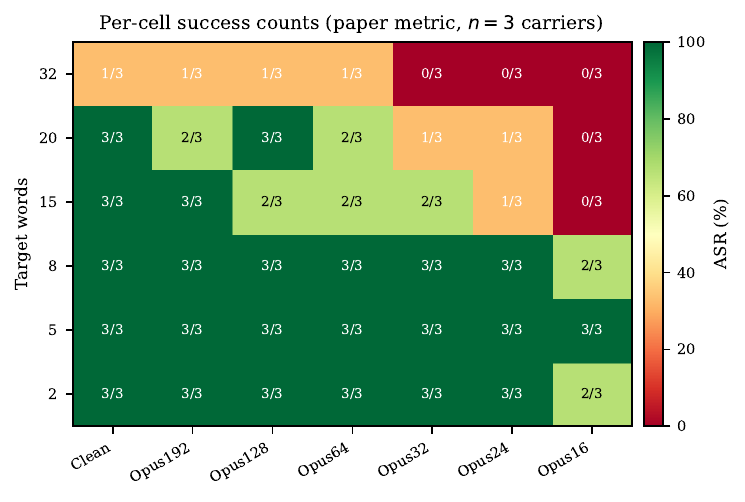}
    \caption{Success counts (out of 3 carriers) for Qwen2-Audio at $\epsilon{=}1.0$ across target word count and Opus bitrate. The attack saturates at $3/3$ for $w{\leq}8$ at all bitrates and degrades gradually at $w{=}15$--$20$ before collapsing at $w{=}32$, where PGD loss increases ${\sim}5\times$ (from 0.011 to 0.052).}
    \label{fig:capacity_cliff}
\end{figure}

 We sweep target length from 2 to 32 words on S3a music carriers (Qwen2-Audio, $\epsilon{=}1.0$), fixing carrier duration at 25\,s across all conditions for a fair comparison. The attack saturates for targets up to 8 words across all Opus bitrates $\geq$24\,kbps, degrades gradually at 15--20 words, and collapses beyond 32 words where PGD loss increases ${\sim}5\times$ (Figure~\ref{fig:capacity_cliff}). The Opus 16\,kbps column fails earliest in the partially-saturated region, suggesting that bitrate rather than word count is the proximate failure mode within $w{\leq}20$, though the small sample ($n{=}3$) limits confidence. All targets used in the main results (S1 banking commands, S2 interview verdicts, S3 single-clause verdicts) fall within the reliable $w{\leq}20$ regime.

\newpage
\section{Additional Figures}
\label{sec:additional_figures}

\subsection{Decoder Jacobian Analysis}
\label{subsec:decoder_heatmap}

\begin{figure}[H]
    \centering
    \includegraphics[width=0.7\linewidth]{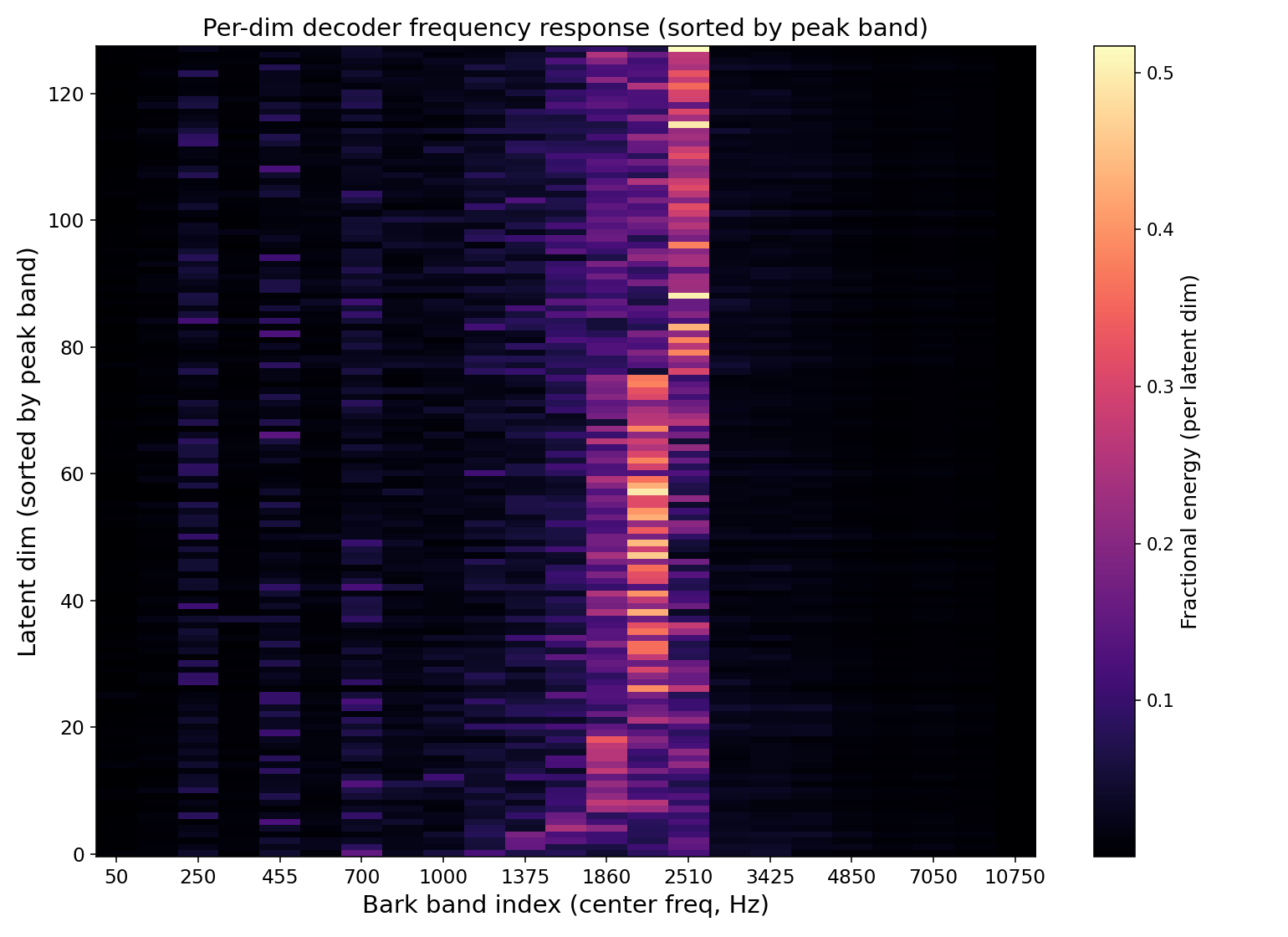}
    \caption{\textbf{EnCodec decoder energy by latent dimension
    and Bark band.} Each row is one of the 128 latent dimensions;
    color indicates fractional output energy in each Bark band,
    computed from the decoder Jacobian $\partial D / \partial z$.
    All dimensions peak in bands 12--14
    (${\approx}$1.8--2.5\,kHz) with negligible energy above
    4\,kHz, showing that the decoder has no basis function
    pointing at the high band. A latent-space perturbation is
    therefore structurally confined to the low-frequency region
    regardless of the adversarial objective.}
    \label{fig:decoder_heatmap}
\end{figure}
Figure~\ref{fig:decoder_heatmap} shows the per-dimension frequency
response of the EnCodec decoder, computed from the Jacobian
$\partial D / \partial \mathbf{z}$. Each row corresponds to one of the 128
latent dimensions, sorted by peak Bark band. The dominant energy for every
dimension falls in Bark bands 12--14 ($\approx$1.8--2.5\,kHz), with a
hard cutoff above 4\,kHz. This confirms that the decoder's convolutional
architecture, trained with multi-resolution STFT losses, has learned basis
functions concentrated in the perceptually most relevant frequency range,
providing no representational capacity for high-frequency perturbations.

Figure~\ref{fig:three_traces} compares three perturbation sources at
matched norm to disentangle the decoder's structural contribution from the
adversarial loss. Source~(A), the Jacobian-derived spectral envelope, is computed by passing a unit perturbation through each of the $D \times F = 128\times 3 = 384$ entries of the latent tensor independently and
aggregating the resulting waveform energy per Bark band. Source~(B), random latent draws at matched scale, has no adversarial objective. Source~(C) is the actual adversarial $\boldsymbol{\delta}$.
Sources~(A) and~(B) overlay band-for-band, both placing approximately
92\% of energy below 4\,kHz, confirming that the sub-4\,kHz confinement
is a property of the decoder parameterization alone. Source~(C) shares
this confinement but further concentrates into the sub-400\,Hz region
(42\% vs.\ approximately 21\% for the non-adversarial sources), where
Opus and MP3 allocate the most bits. The adversarial loss therefore does
not create the low-frequency bias but refines it, shifting energy into the
narrow band the codec preserves most faithfully.

\newpage
\subsection{Speech vs. Music Carriers Analysis}
\label{subsec:speech_vs_music_per_bark}
\begin{figure}[H]
    \centering
    \includegraphics[width=0.98\linewidth]{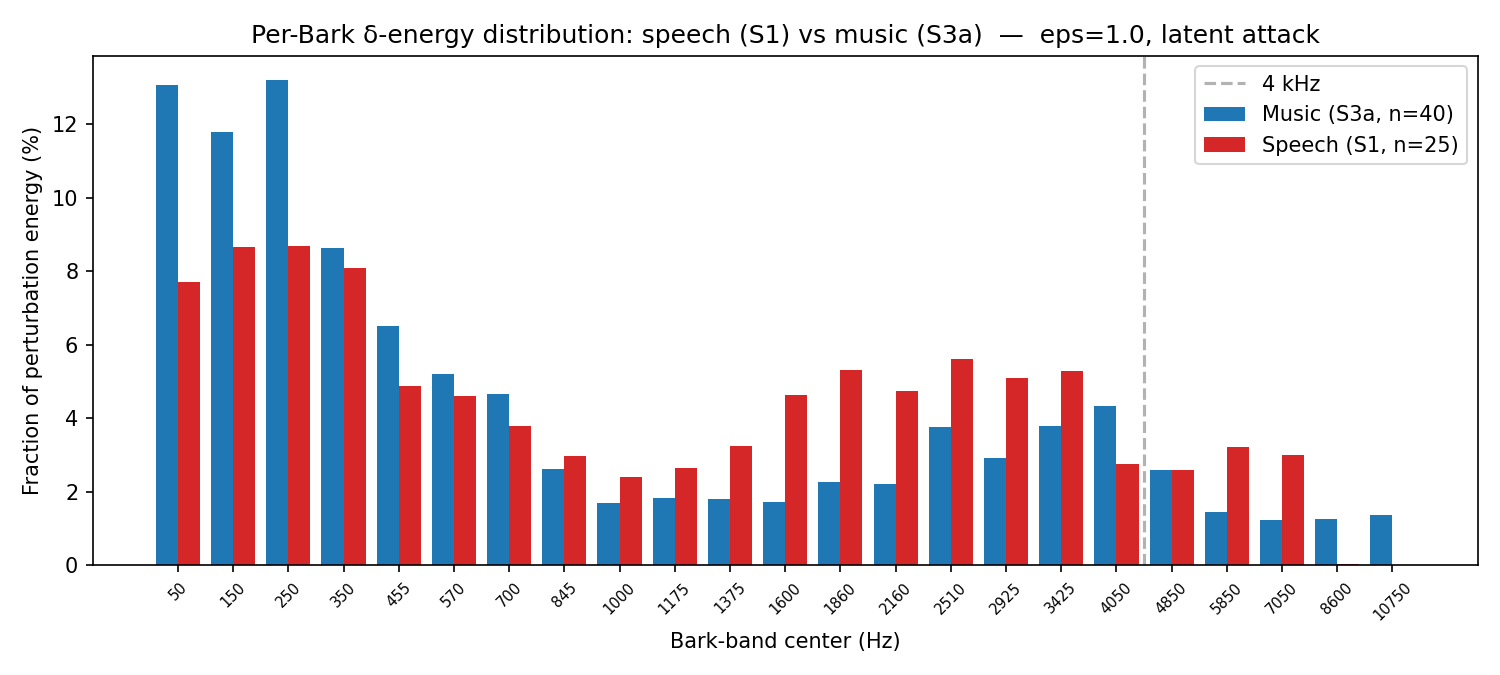}
    \caption{Per-Bark fractional perturbation energy on speech vs.\ music carriers. Music concentrates 1.4$\times$ more $\delta$ energy below 400\,Hz than speech; above 4\,kHz the two distributions are statistically indistinguishable.}
    \label{fig:bark_speech_vs_music}
\end{figure}
The same bit-allocation mechanism explains the S1/S3 gap under AAC-LC. Attack on a music carrier deposits 1.4$\times$ more $\delta$ energy below 400\,Hz than on a speech carrier (Figure~\ref{fig:bark_speech_vs_music}), because music's broadband bass content raises AAC-LC's psychoacoustic masking threshold in that region and the codec preserves whatever sits under the masker. Speech carriers provide no broadband bass masker, so perturbation is forced into the 400\,Hz--4\,kHz formant region where AAC-LC's masking threshold is steepest and $\delta$ is quantized away. Above 4\,kHz the two carrier types are indistinguishable (12.2\% vs.\ 11.6\%), confirming the gap is content-driven through bit allocation, not high-frequency loss.



\subsection{Perturbation Survival}
\label{sec:perturbation_survival}

\begin{figure}[H]
    \centering
    \includegraphics[width=\linewidth]{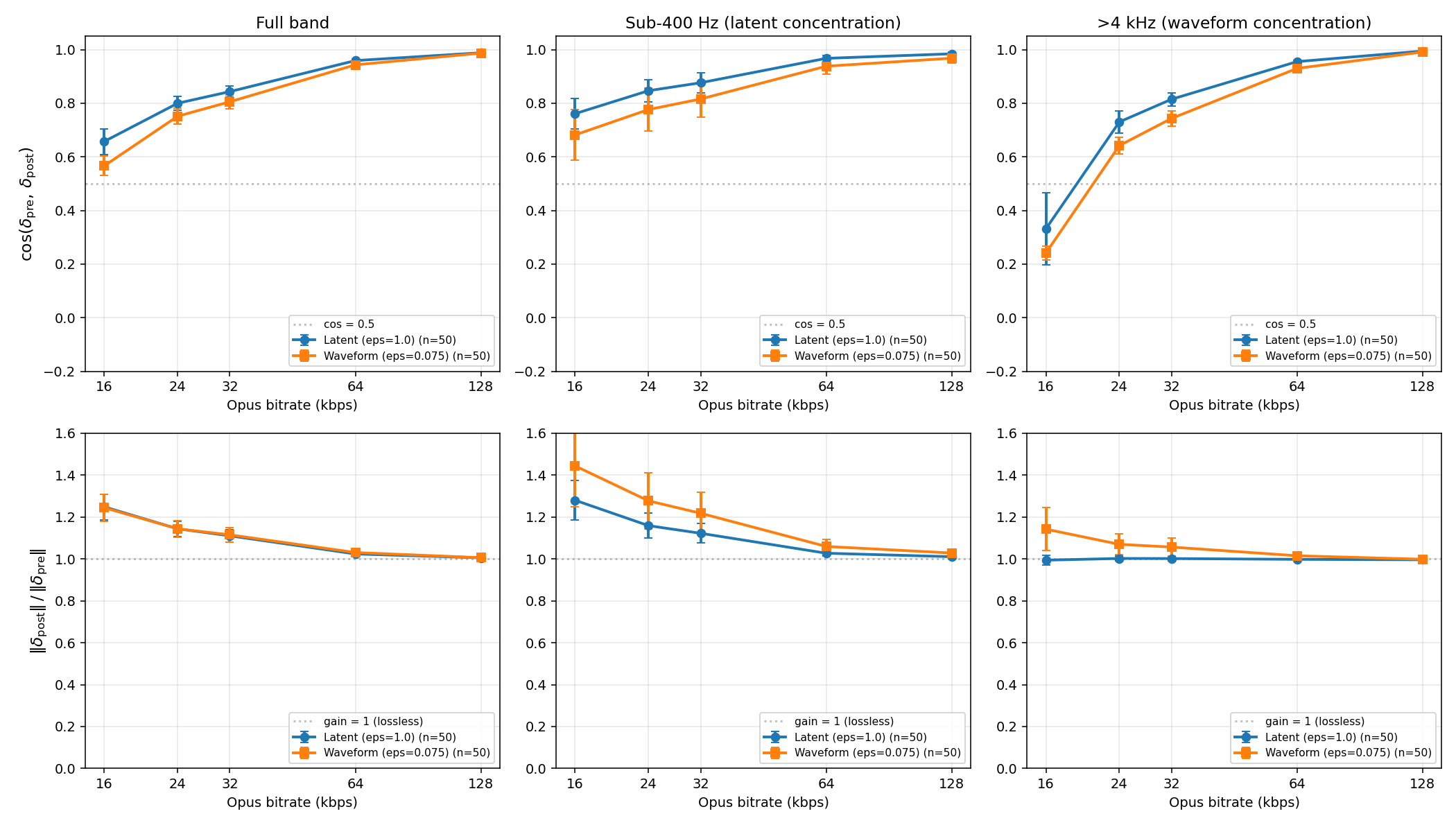}
    \caption{Perturbation survival through Opus at 16--128\,kbps. \textbf{Top}: cosine similarity between pre- and post-codec perturbations. \textbf{Bottom}: magnitude ratio $\|\delta_{\mathrm{post}}\| / \|\delta_{\mathrm{pre}}\|$. Columns split by frequency region. The two attacks are preserved identically within each band.}
    \label{fig:cosine_survival}
\end{figure}

\subsection{Perturbation Spectral Placement}
\label{sec:per_spectral_placement}
\begin{figure}[H]
    \centering
    \includegraphics[width=\linewidth]{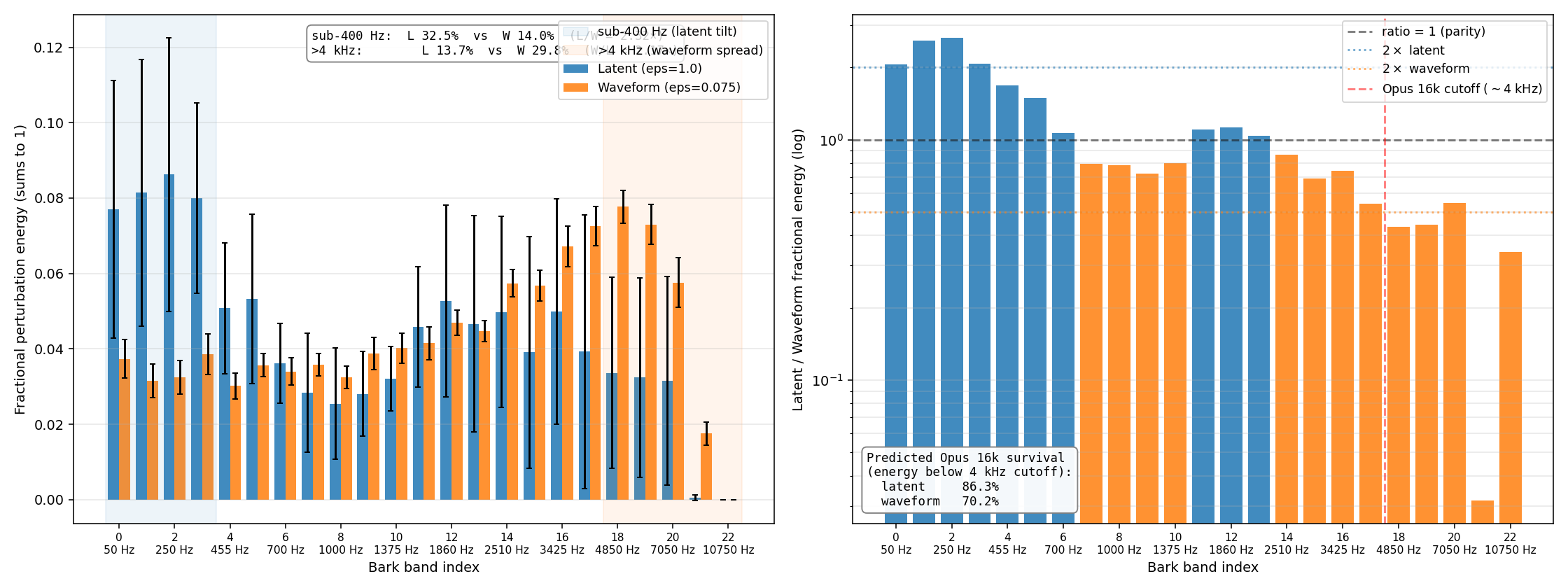}
    \caption{Spectral placement of latent vs.\ waveform perturbations at matched SNR (${\approx}$5.8\,dB). \textbf{Left}: fractional perturbation energy per Bark band. The latent attack concentrates 33.2\% of energy below 400\,Hz; the waveform attack places 29.9\% above 4\,kHz (Spearman $\rho = 0.017$). \textbf{Right}: per-band energy ratio (latent $\div$ waveform, log scale) with the Opus 16\,kbps narrowband cutoff at ${\sim}$4\,kHz.}
    \label{fig:bark_band}
\end{figure}

\section{Additional Analysis}
\label{sec:additional_analysis}





\subsection{Encoder Residual Analysis}
\label{subsec:encoder_residual}
\begin{table}[H]
\centering
\caption{\textbf{Encoder embedding distortion under AAC-LC ($\epsilon{=}1.0$).} $R$ measures the AAC-induced encoder drift normalized by the useful attack displacement. Higher $R$ indicates more attack-relevant information is lost to AAC compression.}
\label{tab:enc_residual_aac}
\begin{tabular}{llcccc}
\toprule
 & & \multicolumn{4}{c}{\textbf{$R$ at AAC-LC bitrate}} \\
\cmidrule(lr){3-6}
\textbf{Model} & \textbf{Carrier} & \textbf{64k} & \textbf{96k} & \textbf{128k} & \textbf{192k} \\
\midrule
\multirow{2}{*}{Qwen2-Audio} & Speech (S1) & \textbf{0.141} & \textbf{0.103} & \textbf{0.103} & \textbf{0.103} \\
 & Music (S3a) & 0.088 & 0.060 & 0.060 & 0.060 \\
\midrule
\multirow{2}{*}{Qwen2.5-Omni} & Speech (S1) & 0.093 & 0.072 & 0.072 & 0.072 \\
 & Music (S3a) & 0.066 & 0.047 & 0.047 & 0.047 \\
\bottomrule
\end{tabular}
\end{table}

To quantify how much attack-relevant information AAC-LC destroys at the encoder level, we define the normalized encoder drift:
\begin{equation}
R = \frac{\lVert h_M(\mathrm{AAC}_b(x_{\mathrm{atk}})) - h_M(x_{\mathrm{atk}}) \rVert}{\lVert h_M(x_{\mathrm{atk}}) - h_M(x_{\mathrm{clean}}) \rVert}
\end{equation}
where $h_M$ is the audio encoder of model $M$, $x_{\mathrm{atk}}$ is the adversarial waveform, and $x_{\mathrm{clean}}$ is the unperturbed carrier. The numerator measures how much AAC compression shifts the encoder's representation of the adversarial audio. The denominator measures the total displacement the attack induces relative to clean audio. When $R$ is small, AAC distortion is negligible compared to the attack signal and the injection survives. When $R$ is large, AAC erases a substantial fraction of the attack displacement and ASR drops.

Table~\ref{tab:enc_residual_aac} reports $R$ for two models across speech and music carriers. The only cell where the attack fails (Qwen2-Audio, speech, S1) is also the only cell with $R > 0.09$. Music carriers consistently yield lower $R$ than speech carriers for both models, consistent with the bit-allocation analysis in \S~\ref{sec:discussion}: music's broadband energy raises the masking threshold, so AAC preserves more of the perturbation. AF3 is omitted due to its incompatible runtime environment for intermediate embedding extraction.

On S1 with AAC-LC, Qwen2.5-Omni retains 54--66\% ASR while
Qwen2-Audio and AF3 collapse to $\leq$2\%
(Tables~\ref{tab:cross_s1_wilson} and~\ref{tab:cross_s3_wilson}).
On S3, all three models retain substantial AAC-LC performance, so
the failure is specific to speech carriers on certain architectures.
The encoder-residual metric $R$ defined above explains the split:
working cells have $R \leq 0.09$, while the failing S1 cells for
Qwen2-Audio and AF3 exceed this threshold. Qwen2.5-Omni stays
below it on the same setting, possibly reflecting its broader audio
training distribution.


\section{Waveform Baseline}
\label{sec:waveform_baseline}

To isolate the contribution of the latent-space parameterization, we compare \texttt{CodecAttack} against a codec-robust \emph{waveform}-space attack that uses the same training recipe as Section~\ref{sec:methodology}.
Compared to the original Carlini--Wagner audio attack~\citep{carlini2018audioadversarialexamplestargeted}, this baseline adds explicit codec robustness: it inherits our multi-bitrate Opus EoT schedule, the BPDA proxy of Equation~\ref{eq:opus_ste}, the two-stage warmup--harden schedule, and the same $S{=}1000$-step Adam optimizer.

\paragraph{Formulation.}
Given carrier $\mathbf{x}$ and target $y^*$, we optimize a waveform-domain perturbation $\boldsymbol{\delta}'\in\mathbb{R}^T$ directly on the 24\,kHz carrier, bypassing the EnCodec encoder/decoder entirely:
\begin{equation}
\label{eq:waveform_attack_clean}
\min_{\boldsymbol{\delta}'} \; \mathcal{L}_{\text{CE}}\!\left(\, f\!\left(R_{16k}(\mathbf{x} + \boldsymbol{\delta}')\right),\; y^* \,\right)
\quad \text{s.t.} \quad \|\boldsymbol{\delta}'\|_\infty \leq \epsilon'.
\end{equation}
The codec-robust variant applies the Opus BPDA proxy $C_{b_t}$ (Equation~\ref{eq:opus_ste}) directly to the perturbed waveform:
\begin{equation}
\label{eq:waveform_codec_robust_loss}
\mathcal{L}^{(t)}_{\text{wave-codec}} \;=\; \mathcal{L}_{\text{CE}}\!\left(\, f\!\left(\, R_{16k}\!\left(\, C_{b_t}(\mathbf{x} + \boldsymbol{\delta}')\right)\right),\; y^* \,\right),
\quad b_t \sim \mathrm{Uniform}(\mathcal{B}).
\end{equation}
Algorithm~\ref{alg:codec_robust} applies unchanged with $\mathbf{z}{+}\boldsymbol{\delta}$ replaced by $\mathbf{x}{+}\boldsymbol{\delta}'$ and $D(\cdot)$ dropped; warmup ratio, step count, EoT grid, and projection schedule are identical.
Adam learning rate is $\alpha'{=}\epsilon'/5$ (the standard waveform-attack choice; learning rate must scale with budget because the waveform $\ell_\infty$ ball is much smaller than the latent one).

\paragraph{SNR-matched budgets.}
Latent and waveform $\ell_\infty$ budgets live in different numeric spaces---$\epsilon$ bounds EnCodec's continuous latent ($\mathbb{R}^{D\times F}$, roughly pre-RVQ magnitudes), whereas $\epsilon'$ bounds PCM amplitudes in $[-1,1]$---so absolute values are not directly comparable.
We therefore match the two attacks by \emph{carrier SNR}: for each reported pair $(\epsilon,\epsilon')$ the clean-channel SNR of the adversarial waveform differs by less than $1$\,dB between the two parameterizations.

\paragraph{Purpose.}
Because the waveform baseline uses the same codec-robust training recipe, codec-EoT schedule, optimizer, and SNR-matched budget as \texttt{CodecAttack}, any remaining gap between the two is attributable to the parameterization alone---not to better training, more iterations, or a laxer perturbation budget.

\newpage
\section{Additional Results}
\subsection{Latent vs. Waveform Attack}
\label{appendix:additional_latent_vs_waveform}
\begin{table}[H]
    \centering
    \small
    \caption{\textbf{Latent vs.\ waveform at low bitrates.} Qwen2-Audio, S1 scenario, $n{=}50$ per attack, identical Opus EoT training, matched SNR (${\approx}$5.8\,dB). ASR (\%). \colorbox{red!10}{Shaded}: higher ASR per cell pair.}
    \label{tab:s1_latent_vs_waveform_low_bitrate}
    \begin{tabular}{lcccccc}
    \toprule
     & \multicolumn{2}{c}{\textbf{Opus}} & \multicolumn{2}{c}{\textbf{MP3 (held-out)}} & \multicolumn{2}{c}{\textbf{AAC-LC (held-out)}} \\
    \cmidrule(lr){2-3} \cmidrule(lr){4-5} \cmidrule(lr){6-7}
    \textbf{Bitrate} & \textbf{Latent} & \textbf{Waveform} & \textbf{Latent} & \textbf{Waveform} & \textbf{Latent} & \textbf{Waveform} \\
    \midrule
    16\,kbps  & \cellcolor{red!10}10.0 & 2.0  & 0.0 & 0.0 & 0.0 & 0.0 \\
    24\,kbps  & \cellcolor{red!10}44.0 & 18.0 & 0.0 & 0.0 & 0.0 & 0.0 \\
    32\,kbps  & \cellcolor{red!10}50.0 & 16.0 & 8.0 & \cellcolor{red!10}10.0 & 0.0 & 0.0 \\
    \bottomrule
    \end{tabular}
\end{table}

\subsection{S1: Financial Voice Agent}
\label{appendix:S1_Wilson}
\begin{table}[H]
    \centering
    \scriptsize
    \setlength{\tabcolsep}{3pt}
    \caption{\textbf{S1 finance voice-agent.} ASR (\%) across all $\epsilon$ values and the full codec evaluation grid.}
    \label{tab:cross_s1_wilson}
    \resizebox{\textwidth}{!}{
    \begin{tabular}{ll c cccccc cccc cccc}
    \toprule
     & & & \multicolumn{6}{c}{\textbf{Opus}} & \multicolumn{4}{c}{\textbf{MP3 (held-out)}} & \multicolumn{4}{c}{\textbf{AAC-LC (held-out)}} \\
    \cmidrule(lr){4-9} \cmidrule(lr){10-13} \cmidrule(lr){14-17}
    \textbf{Model} & \textbf{$\epsilon$} & \textbf{Clean} & \textbf{16k} & \textbf{24k} & \textbf{32k} & \textbf{64k} & \textbf{128k} & \textbf{192k} & \textbf{64k} & \textbf{96k} & \textbf{128k} & \textbf{192k} & \textbf{64k} & \textbf{96k} & \textbf{128k} & \textbf{192k} \\
    \midrule
    \multirow{3}{*}{Qwen2-Audio} & 0.5 & 58.0 & 0.0 & 16.0 & 22.0 & 48.0 & 60.0 & 58.0 & 40.0 & 58.0 & 60.0 & 60.0 & 0.0 & 0.0 & 0.0 & 0.0 \\
                                 & 1.0 & 88.0 & 10.0 & 44.0 & 50.0 & 80.0 & 88.0 & 88.0 & 74.0 & 84.0 & 88.0 & 90.0 & 2.0 & 2.0 & 2.0 & 2.0 \\
                                 & 1.5 & 92.0 & 36.0 & 62.0 & 72.0 & 92.0 & 92.0 & 92.0 & 78.0 & 94.0 & 94.0 & 92.0 & 0.0 & 0.0 & 0.0 & 0.0 \\
    \cmidrule(lr){1-17}
    \multirow{3}{*}{Qwen2.5-Omni} & 0.5 & 30.0 & 4.0 & 16.0 & 20.0 & 30.0 & 28.0 & 30.0 & 30.0 & 30.0 & 30.0 & 30.0 & 12.0 & 16.0 & 18.0 & 20.0 \\
                                  & 1.0 & 82.0 & 36.0 & 56.0 & 62.0 & 76.0 & 82.0 & 82.0 & 80.0 & 80.0 & 82.0 & 82.0 & 54.0 & 66.0 & 66.0 & 66.0 \\
                                  & 1.5 & 90.0 & 44.0 & 76.0 & 80.0 & 88.0 & 90.0 & 90.0 & 88.0 & 90.0 & 90.0 & 90.0 & 80.0 & 86.0 & 86.0 & 86.0 \\
    \cmidrule(lr){1-17}
    \multirow{3}{*}{AF3} & 0.5 & 6.0 & 2.0 & 2.0 & 2.0 & 4.0 & 6.0 & 6.0 & 2.0 & 6.0 & 6.0 & 6.0 & 0.0 & 0.0 & 0.0 & 0.0 \\
                         & 1.0 & 44.0 & 2.0 & 12.0 & 14.0 & 40.0 & 48.0 & 44.0 & 24.0 & 44.0 & 38.0 & 40.0 & 0.0 & 0.0 & 0.0 & 0.0 \\
                         & 1.5 & 68.0 & 2.0 & 22.0 & 40.0 & 54.0 & 68.0 & 66.0 & 38.0 & 64.0 & 68.0 & 66.0 & 0.0 & 0.0 & 0.0 & 0.0 \\
    \bottomrule
    \end{tabular}
    }
\end{table}

\subsection{S2: Interview Screening Scenario}
\label{sec:s2_full_results}

\begin{table}[H]
    \centering
    \scriptsize
    \setlength{\tabcolsep}{3pt}
    \caption{\textbf{S2 interview screening (English carriers).} ASR (\%) across all $\epsilon$ values and the full codec evaluation grid.}
    \label{tab:cross_s2}
    \resizebox{\textwidth}{!}{
    \begin{tabular}{ll c cccccc cccc cccc}
    \toprule
     & & & \multicolumn{6}{c}{\textbf{Opus}} & \multicolumn{4}{c}{\textbf{MP3 (held-out)}} & \multicolumn{4}{c}{\textbf{AAC-LC (held-out)}} \\
    \cmidrule(lr){4-9} \cmidrule(lr){10-13} \cmidrule(lr){14-17}
    \textbf{Model} & \textbf{$\epsilon$} & \textbf{Clean} & \textbf{16k} & \textbf{24k} & \textbf{32k} & \textbf{64k} & \textbf{128k} & \textbf{192k} & \textbf{64k} & \textbf{96k} & \textbf{128k} & \textbf{192k} & \textbf{64k} & \textbf{96k} & \textbf{128k} & \textbf{192k} \\
    \midrule
    \multirow{3}{*}{Qwen2-Audio} & 0.5 & 36.0 & 4.0 & 4.0 & 4.0 & 8.0 & 40.0 & 36.0 & 8.0 & 36.0 & 36.0 & 28.0 & 4.0 & 4.0 & 4.0 & 4.0 \\
                                 & 1.0 & 92.0 & 4.0 & 20.0 & 32.0 & 88.0 & 88.0 & 92.0 & 76.0 & 92.0 & 92.0 & 96.0 & 4.0 & 8.0 & 8.0 & 8.0 \\
                                 & 1.5 & 88.0 & 4.0 & 40.0 & 76.0 & 88.0 & 88.0 & 88.0 & 76.0 & 92.0 & 84.0 & 88.0 & 0.0 & 0.0 & 0.0 & 0.0 \\
    \cmidrule(lr){1-17}
    \multirow{3}{*}{Qwen2.5-Omni} & 0.5 & 44.0 & 24.0 & 36.0 & 32.0 & 44.0 & 44.0 & 44.0 & 44.0 & 44.0 & 44.0 & 44.0 & 36.0 & 40.0 & 40.0 & 40.0 \\
                                  & 1.0 & 84.0 & 48.0 & 72.0 & 76.0 & 84.0 & 84.0 & 84.0 & 84.0 & 84.0 & 84.0 & 84.0 & 80.0 & 80.0 & 76.0 & 76.0 \\
                                  & 1.5 & 88.0 & 48.0 & 76.0 & 88.0 & 88.0 & 88.0 & 88.0 & 88.0 & 88.0 & 88.0 & 88.0 & 76.0 & 80.0 & 80.0 & 84.0 \\
    \cmidrule(lr){1-17}
    \multirow{3}{*}{AF3} & 0.5 & 4.0 & 0.0 & 4.0 & 4.0 & 4.0 & 4.0 & 4.0 & 4.0 & 4.0 & 4.0 & 4.0 & 0.0 & 0.0 & 0.0 & 0.0 \\
                         & 1.0 & 36.0 & 0.0 & 8.0 & 12.0 & 24.0 & 36.0 & 36.0 & 16.0 & 32.0 & 36.0 & 32.0 & 0.0 & 0.0 & 0.0 & 0.0 \\
                         & 1.5 & 60.0 & 0.0 & 12.0 & 24.0 & 52.0 & 60.0 & 56.0 & 40.0 & 60.0 & 60.0 & 60.0 & 0.0 & 0.0 & 0.0 & 0.0 \\
    \bottomrule
    \end{tabular}
    }
\end{table}
\begin{table}[H]
    \centering
    \scriptsize
    \setlength{\tabcolsep}{3pt}
    \caption{\textbf{S2 on Mandarin.} ASR (\%) across all $\epsilon$ values and the full codec evaluation grid. Switching the carrier language drops Qwen2-Audio ASR at $\epsilon{=}1.0$ from 92\% (English, Table~\ref{tab:cross_s2}) to 29\%.}
    \label{tab:cross_s2_zh}
    \resizebox{\textwidth}{!}{
    \begin{tabular}{ll c cccccc cccc cccc}
    \toprule
     & & & \multicolumn{6}{c}{\textbf{Opus}} & \multicolumn{4}{c}{\textbf{MP3 (held-out)}} & \multicolumn{4}{c}{\textbf{AAC-LC (held-out)}} \\
    \cmidrule(lr){4-9} \cmidrule(lr){10-13} \cmidrule(lr){14-17}
    \textbf{Model} & \textbf{$\epsilon$} & \textbf{Clean} & \textbf{16k} & \textbf{24k} & \textbf{32k} & \textbf{64k} & \textbf{128k} & \textbf{192k} & \textbf{64k} & \textbf{96k} & \textbf{128k} & \textbf{192k} & \textbf{64k} & \textbf{96k} & \textbf{128k} & \textbf{192k} \\
    \midrule
    \multirow{3}{*}{Qwen2-Audio} & 0.5 & 0.0 & 0.0 & 0.0 & 0.0 & 0.0 & 0.0 & 0.0 & 0.0 & 0.0 & 0.0 & 0.0 & 0.0 & 0.0 & 0.0 & 0.0 \\
                                 & 1.0 & 29.2 & 0.0 & 4.2 & 12.5 & 20.8 & 29.2 & 29.2 & 16.7 & 25.0 & 29.2 & 25.0 & 4.2 & 20.8 & 20.8 & 20.8 \\
                                 & 1.5 & 41.7 & 12.5 & 16.7 & 20.8 & 33.3 & 41.7 & 45.8 & 33.3 & 41.7 & 41.7 & 41.7 & 16.7 & 33.3 & 33.3 & 29.2 \\
    \cmidrule(lr){1-17}
    \multirow{3}{*}{Qwen2.5-Omni} & 0.5 & 0.0 & 0.0 & 0.0 & 0.0 & 0.0 & 0.0 & 0.0 & 0.0 & 0.0 & 0.0 & 0.0 & 0.0 & 0.0 & 0.0 & 0.0 \\
                                  & 1.0 & 20.8 & 4.2 & 12.5 & 16.7 & 20.8 & 20.8 & 20.8 & 16.7 & 20.8 & 20.8 & 20.8 & 16.7 & 20.8 & 20.8 & 20.8 \\
                                  & 1.5 & 33.3 & 4.2 & 12.5 & 12.5 & 20.8 & 33.3 & 29.2 & 20.8 & 33.3 & 33.3 & 33.3 & 8.3 & 25.0 & 25.0 & 25.0 \\
    \cmidrule(lr){1-17}
    \multirow{3}{*}{AF3} & 0.5 & 4.2 & 0.0 & 0.0 & 0.0 & 4.2 & 4.2 & 4.2 & 4.2 & 4.2 & 4.2 & 4.2 & 0.0 & 0.0 & 0.0 & 0.0 \\
                         & 1.0 & 20.8 & 0.0 & 4.2 & 4.2 & 16.7 & 25.0 & 20.8 & 12.5 & 25.0 & 20.8 & 25.0 & 0.0 & 4.2 & 8.3 & 12.5 \\
                         & 1.5 & 41.7 & 0.0 & 0.0 & 4.2 & 25.0 & 41.7 & 41.7 & 12.5 & 41.7 & 41.7 & 41.7 & 0.0 & 8.3 & 8.3 & 8.3 \\
    \bottomrule
    \end{tabular}
    }
\end{table}

\subsection{S3: Music Industry Bypass}
\label{appendix:S3_Wilson}
\begin{table}[H]
    \centering
    \scriptsize
    \setlength{\tabcolsep}{3pt}
    \caption{\textbf{S3 music-industry bypass} (S3a: AI-detection; S3b: copyright). ASR (\%) across all $\epsilon$ values and the full codec evaluation grid.}
    \label{tab:cross_s3_wilson}
    \resizebox{\textwidth}{!}{
    \begin{tabular}{lll c cccccc cccc cccc}
    \toprule
     & & & & \multicolumn{6}{c}{\textbf{Opus}} & \multicolumn{4}{c}{\textbf{MP3 (held-out)}} & \multicolumn{4}{c}{\textbf{AAC-LC (held-out)}} \\
    \cmidrule(lr){5-10} \cmidrule(lr){11-14} \cmidrule(lr){15-18}
    \textbf{Scenario} & \textbf{Model} & \textbf{$\epsilon$} & \textbf{Clean} & \textbf{16k} & \textbf{24k} & \textbf{32k} & \textbf{64k} & \textbf{128k} & \textbf{192k} & \textbf{64k} & \textbf{96k} & \textbf{128k} & \textbf{192k} & \textbf{64k} & \textbf{96k} & \textbf{128k} & \textbf{192k} \\
    \midrule
    \multirow{9}{*}{S3a}
      & \multirow{3}{*}{Qwen2-Audio} & 0.5 & 87.5 & 0.0 & 10.0 & 15.0 & 70.0 & 92.5 & 87.5 & 55.0 & 82.5 & 87.5 & 87.5 & 15.0 & 65.0 & 65.0 & 72.5 \\
      &  & 1.0 & 97.5 & 17.5 & 40.0 & 60.0 & 87.5 & 100.0 & 97.5 & 90.0 & 92.5 & 95.0 & 97.5 & 47.5 & 82.5 & 85.0 & 87.5 \\
      &  & 1.5 & 97.5 & 42.5 & 75.0 & 92.5 & 100.0 & 100.0 & 100.0 & 92.5 & 100.0 & 97.5 & 100.0 & 77.5 & 90.0 & 92.5 & 92.5 \\
      \cmidrule(lr){2-18}
      & \multirow{3}{*}{Qwen2.5-Omni} & 0.5 & 60.0 & 0.0 & 0.0 & 0.0 & 12.5 & 60.0 & 60.0 & 22.5 & 57.5 & 60.0 & 60.0 & 2.5 & 45.0 & 47.5 & 47.5 \\
      &  & 1.0 & 95.0 & 22.5 & 35.0 & 47.5 & 87.5 & 92.5 & 92.5 & 72.5 & 82.5 & 85.0 & 85.0 & 42.5 & 65.0 & 65.0 & 65.0 \\
      &  & 1.5 & 100.0 & 30.0 & 55.0 & 72.5 & 97.5 & 100.0 & 100.0 & 92.5 & 100.0 & 100.0 & 100.0 & 62.5 & 75.0 & 77.5 & 77.5 \\
      \cmidrule(lr){2-18}
      & \multirow{3}{*}{AF3} & 0.5 & 65.0 & 0.0 & 5.0 & 10.0 & 52.5 & 65.0 & 65.0 & 41.0 & 56.4 & 56.4 & 56.4 & 0.0 & 25.6 & 25.6 & 28.2 \\
      &  & 1.0 & 97.5 & 2.5 & 30.0 & 57.5 & 95.0 & 97.5 & 95.0 & 82.5 & 95.0 & 95.0 & 95.0 & 22.5 & 60.0 & 65.0 & 65.0 \\
      &  & 1.5 & 97.5 & 12.5 & 45.0 & 82.5 & 95.0 & 95.0 & 92.5 & 82.5 & 95.0 & 92.5 & 95.0 & 40.0 & 65.0 & 65.0 & 67.5 \\
    \midrule
    \multirow{9}{*}{S3b}
      & \multirow{3}{*}{Qwen2-Audio} & 0.5 & 93.3 & 2.2 & 24.4 & 42.2 & 84.4 & 100.0 & 95.6 & 75.6 & 86.7 & 91.1 & 93.3 & 33.3 & 66.7 & 68.9 & 71.1 \\
      &  & 1.0 & 100.0 & 53.3 & 75.6 & 93.3 & 97.8 & 100.0 & 100.0 & 95.6 & 100.0 & 100.0 & 100.0 & 73.3 & 86.7 & 88.9 & 88.9 \\
      &  & 1.5 & 100.0 & 86.7 & 97.8 & 100.0 & 100.0 & 100.0 & 100.0 & 100.0 & 100.0 & 100.0 & 100.0 & 77.8 & 84.4 & 84.4 & 84.4 \\
      \cmidrule(lr){2-18}
      & \multirow{3}{*}{Qwen2.5-Omni} & 0.5 & 75.6 & 0.0 & 2.2 & 4.4 & 35.6 & 75.6 & 75.6 & 40.0 & 73.3 & 73.3 & 75.6 & 8.9 & 46.7 & 48.9 & 48.9 \\
      &  & 1.0 & 100.0 & 28.9 & 46.7 & 64.4 & 93.3 & 100.0 & 100.0 & 84.4 & 100.0 & 100.0 & 100.0 & 55.6 & 82.2 & 84.4 & 82.2 \\
      &  & 1.5 & 100.0 & 42.2 & 71.1 & 84.4 & 95.6 & 100.0 & 97.8 & 93.3 & 97.8 & 97.8 & 97.8 & 68.9 & 86.7 & 88.9 & 86.7 \\
      \cmidrule(lr){2-18}
      & \multirow{3}{*}{AF3} & 0.5 & 97.8 & 4.4 & 17.8 & 26.7 & 88.9 & 100.0 & 97.8 & 62.2 & 93.3 & 97.8 & 97.8 & 22.2 & 60.0 & 64.4 & 68.9 \\
      &  & 1.0 & 100.0 & 20.0 & 57.8 & 88.9 & 95.6 & 100.0 & 97.8 & 86.7 & 97.8 & 97.8 & 97.8 & 53.3 & 73.3 & 77.8 & 77.8 \\
      &  & 1.5 & 97.8 & 60.0 & 84.4 & 95.6 & 97.8 & 100.0 & 100.0 & 91.1 & 88.9 & 93.3 & 91.1 & 57.8 & 80.0 & 77.8 & 80.0 \\
    \bottomrule
    \end{tabular}
    }
\end{table}

\newpage
\section{Audio Quality Results}
\label{sec:audio_quality}
\begin{table}[H]
\centering
\caption{\textbf{Audio quality of adversarial carriers}, averaged across the three victim models per scenario; quality degrades smoothly with $\epsilon$, the cost of placing energy in codec-preserved bands. $\uparrow$/$\downarrow$: higher/lower is better.}
\vspace{5pt}
\label{tab:audio_quality}
\resizebox{\textwidth}{!}{%
\begin{tabular}{llrrrrrr}
\toprule
\textbf{Scenario} & $\boldsymbol{\epsilon}$ & \textbf{SNR (dB)} & \textbf{SI-SDR (dB)} & \textbf{PESQ-WB}~$\uparrow$ & \textbf{STOI}~$\uparrow$ & \textbf{ESTOI}~$\uparrow$ & \textbf{LSD (dB)}~$\downarrow$ \\
\midrule
\multirow{3}{*}{S1 (Finance)} 
  & 0.5 & 8.52 $\pm$ 2.45 & 8.44 $\pm$ 2.82 & 2.63 $\pm$ 0.26 & 0.959 $\pm$ 0.007 & 0.914 $\pm$ 0.015 & 34.07 $\pm$ 4.32 \\
  & 1.0 & 6.12 $\pm$ 1.99 & 5.93 $\pm$ 2.34 & 1.69 $\pm$ 0.20 & 0.904 $\pm$ 0.017 & 0.806 $\pm$ 0.035 & 37.18 $\pm$ 4.53 \\
  & 1.5 & 3.74 $\pm$ 1.81 & 3.41 $\pm$ 2.05 & 1.33 $\pm$ 0.14 & 0.847 $\pm$ 0.025 & 0.700 $\pm$ 0.048 & 39.96 $\pm$ 4.66 \\
\midrule
\multirow{3}{*}{S2 (Interview, EN)} 
  & 0.5 & 9.73 $\pm$ 2.01 & 9.90 $\pm$ 2.31 & 2.69 $\pm$ 0.23 & 0.952 $\pm$ 0.024 & 0.915 $\pm$ 0.034 & 32.61 $\pm$ 11.84 \\
  & 1.0 & 7.25 $\pm$ 1.61 & 7.33 $\pm$ 1.87 & 1.73 $\pm$ 0.20 & 0.901 $\pm$ 0.035 & 0.823 $\pm$ 0.044 & 35.67 $\pm$ 12.52 \\
  & 1.5 & 4.85 $\pm$ 1.39 & 4.73 $\pm$ 1.69 & 1.31 $\pm$ 0.14 & 0.841 $\pm$ 0.041 & 0.723 $\pm$ 0.057 & 38.52 $\pm$ 12.98 \\
\midrule
\multirow{3}{*}{S3a (AI Detection)} 
  & 0.5 & 5.12 $\pm$ 3.79 & 6.27 $\pm$ 3.06 & 2.06 $\pm$ 0.22 & 0.770 $\pm$ 0.078 & 0.651 $\pm$ 0.080 & 8.19 $\pm$ 0.90 \\
  & 1.0 & 3.94 $\pm$ 3.56 & 4.63 $\pm$ 2.64 & 1.46 $\pm$ 0.13 & 0.668 $\pm$ 0.087 & 0.528 $\pm$ 0.080 & 9.10 $\pm$ 1.19 \\
  & 1.5 & 2.69 $\pm$ 3.29 & 2.86 $\pm$ 2.38 & 1.25 $\pm$ 0.10 & 0.579 $\pm$ 0.091 & 0.430 $\pm$ 0.077 & 10.00 $\pm$ 1.44 \\
\midrule
\multirow{3}{*}{S3b (Copyright)} 
  & 0.5 & 0.64 $\pm$ 7.88 & 4.65 $\pm$ 1.54 & 2.06 $\pm$ 0.44 & 0.701 $\pm$ 0.176 & 0.645 $\pm$ 0.184 & 10.76 $\pm$ 3.01 \\
  & 1.0 & 0.00 $\pm$ 7.65 & 3.60 $\pm$ 1.35 & 1.61 $\pm$ 0.26 & 0.610 $\pm$ 0.201 & 0.547 $\pm$ 0.191 & 11.31 $\pm$ 3.19 \\
  & 1.5 & $-$0.78 $\pm$ 7.48 & 2.34 $\pm$ 1.39 & 1.37 $\pm$ 0.16 & 0.533 $\pm$ 0.211 & 0.463 $\pm$ 0.187 & 11.92 $\pm$ 3.48 \\
\bottomrule
\end{tabular}%
}
\end{table}

Table~\ref{tab:audio_quality} reports audio quality metrics averaged across the three target models per scenario. Quality degrades smoothly with $\epsilon$, reflecting the fundamental tradeoff: codec robustness requires perturbation energy in codec-preserved frequency bands, and that energy is audible.

For speech carriers (S1, S2), PESQ-WB and STOI are the primary indicators. At $\epsilon{=}1.0$ both scenarios retain STOI ${\approx}$0.90, indicating near-full intelligibility despite moderate quality loss (PESQ ${\approx}$1.7). At $\epsilon{=}0.5$ quality approaches transparent (PESQ ${\approx}$2.6, STOI ${\approx}$0.95) but ASR drops correspondingly (see main results). At $\epsilon{=}1.5$ intelligibility remains acceptable (STOI ${\approx}$0.84) at the cost of noticeable distortion. The PESQ-WB drop at $\epsilon{\geq}1.0$ is a direct consequence of the bit-allocation mechanism (\S\ref{sec:discussion}): a codec-robust perturbation must place energy in psychoacoustically relevant speech bands, which is exactly the bands PESQ penalizes. The trade-off is therefore quality versus codec survival; a perturbation with PESQ approaching 4 would, by construction, occupy frequencies the codec strips and would not survive the channel. The intelligibility metric (STOI ${\approx}$0.90) and the downstream ASR confirm that the model still parses the carrier.

For music carriers (S3a, S3b), PESQ-WB and STOI are less informative since these metrics are designed for speech. Log-Spectral Distance (LSD) and SI-SDR better characterize music quality. At $\epsilon{=}1.0$, S3a yields LSD ${\approx}$9.1\,dB and SI-SDR ${\approx}$4.6\,dB, while S3b shows LSD ${\approx}$11.3\,dB and SI-SDR ${\approx}$3.6\,dB. The higher variance on S3b SNR reflects the wide dynamic-range spread across music genres (jazz, classical, calm, Christmas jazz). Perceptually, the adversarial music carriers retain their genre character and are not obviously corrupted to a casual listener.

The $\epsilon$ values in this work are not directly comparable to the perturbation budgets of waveform attacks on transcription models~\citep{carlini2018audioadversarialexamplestargeted, qin2019imperceptiblerobusttargetedadversarial, schonherr2018adversarial}, which bound PCM amplitude and assume clean-channel delivery. Our perturbation occupies a latent space with different dimensionality and dynamic range. The appropriate comparison is on output quality metrics (STOI, PESQ), not raw $\epsilon$ values. On this basis, \texttt{CodecAttack} at $\epsilon{=}1.0$ achieves higher STOI on speech carriers than Kim et al.~\citep{kim2025good} (STOI ${\approx}$0.59--0.70 at waveform $\epsilon{=}0.2$--$0.5$).

\section{Background}
\label{app_sec:background}


\subsection{Neural Audio Codecs} 

Neural audio codecs compress waveforms into compact discrete representations via an encoder--quantizer--decoder pipeline. SoundStream~\citep{zeghidour2021soundstreamendtoendneuralaudio} introduced residual vector quantization (RVQ) for end-to-end neural compression; EnCodec~\citep{defossez2022high} extended this with multi-scale STFT discriminators, achieving high-fidelity reconstruction at bandwidths as low as 1.5\,kbps. Subsequent work improves reconstruction quality~\citep{kumar2023highfidelityaudiocompressionimproved}, disentangles semantic and acoustic information~\citep{zhang2024speechtokenizerunifiedspeechtokenizer}, provides reproducible open-source tooling~\citep{du2023funcodecfundamentalreproducibleintegrable}, and compresses token rates further while maintaining quality~\citep{ji2025wavtokenizerefficientacousticdiscrete}.

\subsection{Audio Large Language Models} 

The dominant recipe for audio understanding pairs a pre-trained audio encoder with a large language model through a lightweight adapter. Whisper~\citep{radford2022robustspeechrecognitionlargescale} provides a widely adopted encoder backbone; subsequent systems build on this foundation to support progressively richer tasks, from audio question answering~\citep{deshmukh2024pengiaudiolanguagemodel, gong2024listenthinkunderstand} to joint speech and non-speech understanding~\citep{tang2024salmonngenerichearingabilities}. Qwen-Audio~\citep{chu2023qwenaudioadvancinguniversalaudio} and Qwen2-Audio~\citep{chu2024qwen2audiotechnicalreport} scale this approach with multi-task pre-training across 30+ datasets and introduce voice-interaction modes with instruction-following capabilities. A parallel line of work replaces continuous encoder features with discrete codec tokens: AudioLM~\citep{borsos2023audiolmlanguagemodelingapproach} models SoundStream tokens autoregressively, VALL-E~\citep{wang2023neuralcodeclanguagemodels} leverages EnCodec tokens for zero-shot TTS, and MusicGen~\citep{copet2024simplecontrollablemusicgeneration} adopts the same tokenization for controllable music generation. Because Audio LLMs accept raw waveforms and produce free-form text or actions, they present a qualitatively broader attack surface than traditional ASR pipelines: an adversary can target arbitrary downstream behaviors (including tool use and jailbreaking) not merely transcription errors.

\section{Full Experimental Setup}
\label{sec:full_setup}

This section provides complete details for the experimental design summarized in \S~\ref{sec:experimental_setup}.

\textbf{Evaluation metric.} We use strict substring match rather than a WER-threshold fallback because the latter scores near-miss outputs unevenly across attack types and inflates apparent ASR. Substring match requires the model to emit the target verbatim, the criterion that corresponds to a downstream tool-calling pipeline executing the injected action. Output is normalized (lower-cased, whitespace-collapsed, punctuation-stripped) before matching.

\textbf{Scenario design.} Each scenario pairs every carrier waveform with an attacker-chosen target command drawn from a category-balanced map so that no single phrasing dominates.
\begin{itemize}
    \item \textbf{S1} uses 25 English banking speech clips paired with targets from five authorization-bypass categories (PIN/auth bypass, wire transfer, account disclosure, fraud-flag bypass, tool-call injection), five phrasings per category, yielding 50 unique targets mapped one-to-one to the carriers.
    \item \textbf{S2} uses 25 English and 24 Mandarin interview speech clips whose content describes underqualified candidates so that any ``Strongly Recommend Advancing'' verdict is attributable to the perturbation, not content merit.
    \item \textbf{S3} uses ${\sim}$25-second music clips at 24\,kHz (EnCodec's native rate) spanning jazz, classical, calm, and Christmas jazz: 40 carriers for S3a (AI-detection spoofing) and 45 for S3b (copyright-classifier spoofing), each paired with a single-clause verdict target.
\end{itemize}
In all cases the adversarial artifact delivered to the victim is a standard 16-bit PCM wav file with no latent-code injection or out-of-band metadata.

\textbf{Codec evaluation grid.} The attack is trained with Opus EoT over $\mathcal{B} = \{16, 24, 32, 64, 128\}$\,kbps and evaluated on $\mathcal{B} \cup \{192\}$\,kbps (held-out bitrate), plus cross-family transfer to MP3 and AAC-LC at $\{64, 96, 128, 192\}$\,kbps each. This grid covers the full bitrate range of real deployment channels and tests whether robustness generalizes beyond the codec family seen during optimization.

\textbf{Perturbation budget.} The bound $\epsilon \in \{0.5, 1.0, 1.5\}$ constrains the $\ell_\infty$ norm of $\boldsymbol{\delta}$ in EnCodec's continuous latent space ($\mathbb{R}^{D \times F}$), not in the waveform domain. These values are not comparable to the perturbation budgets of waveform attacks on transcription models~\citep{carlini2018audioadversarialexamplestargeted, qin2019imperceptiblerobusttargetedadversarial, schonherr2018adversarial}, which bound PCM amplitude and assume clean-channel delivery. Our perturbation occupies a latent space with different dimensionality and dynamic range, so the appropriate comparison is on output quality metrics, not raw $\epsilon$ values.

\textbf{Computing Resource Details.} Each 1000-step optimization takes approximately 8 minutes on a single NVIDIA A100 (80\,GB); the full evaluation across all scenarios, models, and codec compression channels requires approximately 350 GPU-hours.


\end{document}